\documentclass[%
 reprint,
superscriptaddress,
 amsmath,amssymb,
 aps, prx,
longbibliography]{revtex4-2}

\usepackage{graphicx}
\usepackage{dcolumn}
\usepackage{bm}
\usepackage{braket}
\usepackage[caption=false]{subfig}
\usepackage{makecell}
\usepackage{appendix}

\begin{document}

\preprint{APS/123-QED}

\title{\textbf{Self-limiting electrostriction of a single ion in an ultracold polar gas: From mesoscopic ions to crystalline molecular rings } 
}%

\author{Ruiren Shi}
\affiliation{Department of Physics and Astronomy, Stony Brook University, Stony Brook 11794, NY, USA}

\author{Saajid Chowdhury}
\affiliation{Department of Physics and Astronomy, Stony Brook University, Stony Brook, NY 11794, USA}

\author{Leon Karpa}
\affiliation{Leibniz Universität Hannover, Institut für Quantenoptik}

\author{Jes\'us P\'erez-R\'ios}
\email{Contact author: jesus.perezrios@stonybrook.edu}
\affiliation{Department of Physics and Astronomy, Stony Brook University, Stony Brook 11794, NY, USA}

\date{\today}

\begin{abstract}
We investigate the self-assembly of polar molecules around a single ion immersed in an ultracold, dilute two-dimensional molecular gas. The ion aligns and attracts the molecules through charge-dipole interactions, producing a strong electrostrictive accumulation around the impurity, while intermolecular repulsion limits further densification and favors spatially extended configurations. By combining global optimization with diffusion Monte Carlo, we calculate the evaporation energy as a function of the number of molecules bound to the ion. In contrast to conventional charged and van der Waals clusters, the evaporation energy exhibits a plateau-like dependence on cluster size, reflecting the sequential formation of concentric molecular rings. These structures are governed by the topology of the ion’s electric field and by the competition between attractive ion–molecule interactions, repulsive intra-ring interactions, and attractive correlations between neighboring rings, rather than by conventional coordination or icosahedral packing. In the weak-interaction regime, the resulting structures form extended mesoscopic molecular ions, whereas stronger interactions produce increasingly rigid, crystal-like molecular rings. We further analyze their stability against thermal perturbations and the time-dependent ion trap and find that a broad range of clusters remain stable under experimentally relevant conditions. The intermolecular repulsion and dipolar geometry also suppress close-range ion–molecule encounters, suggesting an intrinsic shielding mechanism. Our results establish ion-bound polar-molecule clusters as a distinct class of mesoscopic molecular ions and open a route to studying charged impurities in quantum baths with anisotropic interactions.

\end{abstract}

\maketitle


\section{Introduction}

The advent of ion-atom hybrid systems, which merge ultracold atoms and cold ions, has paved the way for a new era in cold molecular sciences~\cite{Deiss2024}. In cold chemistry, these experimental platforms have proven essential for understanding the fundamental properties underlying charged-neutral interactions, owing to the exquisite control over the internal states and relative energy of the colliding partners~\cite{COTE20166,TomasReview,LOUS202265}. Charged-neutral interactions are dominated by long-range forces and can lead to efficient molecular-ion formation, even in the absence of ordinary bimolecular reaction channels, through termolecular processes involving the surrounding bath~\cite{Krukow2016,Mohammadi2021}. An extreme manifestation of this behavior is the presence of ion-atom Feshbach resonances~\cite{Weckesser2021}, which make it possible to control the ion’s reactivity through termolecular processes by means of an external magnetic field. Another essential characteristic of these systems is that the ion is confined in a time-dependent trap. Although this is often regarded as a complication~\cite{Cetina2012}, the trap can also serve as a control knob, modifying the collision dynamics and potentially favoring the formation of long-lived molecular complexes~\cite{Pinkas2023,Hirzler2023}, in close analogy with the so-called sticky-collision scenario known from ultracold molecular gases~\cite{Sticking,Sticking2}.

On the other hand, a single ion in a neutral bath can be viewed as a charged impurity moving through a polarizable medium. The ion-atom interaction, governed by charge-induced-dipole forces, is stronger and longer-ranged than the typical van der Waals interaction between the atoms in the bath. As a result, the ion interacts simultaneously with several bath atoms, creating a many-body excitation known as an ionic polaron~\cite{astrakharchik2020ionic}. An ionic polaron can be described as a many-body bound state of a single ion with multiple atoms when no atom-atom bound states are accessible; in this regime, the impurity-bath interaction itself establishes the binding mechanism. By contrast, when atom-atom bound states are possible, a single ion can bind many atoms together, forming mesoscopic molecular ions~\cite{Meso3,Meso1}. These are massive molecular ions containing hundreds of atoms, whose stability is dictated by the interplay between few-body physics, long-range interactions, and the surrounding bath.

Alongside these developments, the cooling and control of polar molecules has grown into one of the central pillars of modern atomic, molecular, and optical physics~\cite{UltracoldMolecules,UltracoldChemistry}. A significant effort has been devoted to understanding polar molecules in ultracold environments and, in particular, to controlling their short-range interactions in order to avoid collisional loss and sticky collisions~\cite{Sticking, Sticking2}. Thanks to shielding techniques, including microwave, optical, and static shielding~\cite{Shielding,Shielding_observation,Karman2018,Lassabliere2018,Karam2023,Quemener2010}, it has become possible to suppress short-range losses and to experimentally demonstrate the first molecular Bose-Einstein condensate of polar molecules~\cite{Bigagli2024}. Moreover, polar molecules can now be trapped in optical tweezers~\cite{Loic2019}, opening new possibilities for quantum simulation, quantum information processing, and the engineering of many-body quantum systems~\cite{PRXQuantum.5.020333,applications}.

Motivated by the proposal of Karpa and Dulieu~\cite{Karpa2025} to combine a trapped ion with an ultracold gas of polar molecules, we investigate the stability and structure of a single ion immersed in a polar molecular bath. This setting realizes a new form of quantum matter: a mesoscopic molecular ion stabilized not by conventional short-range chemical bonding or by isotropic long-range interactions that lead to spherical packing, as in van der Waals clusters, but by self-limiting electrostriction. The ion acts as a self-organizing center that aligns the molecules, while the repulsive dipole-dipole interactions within a ring frustrate collapse and favor spatially extended structures. The repulsion can be strong enough that mesoscopic ions evolve into crystalline molecular rings.

\begin{figure*}[t]
    \includegraphics[width=\linewidth]{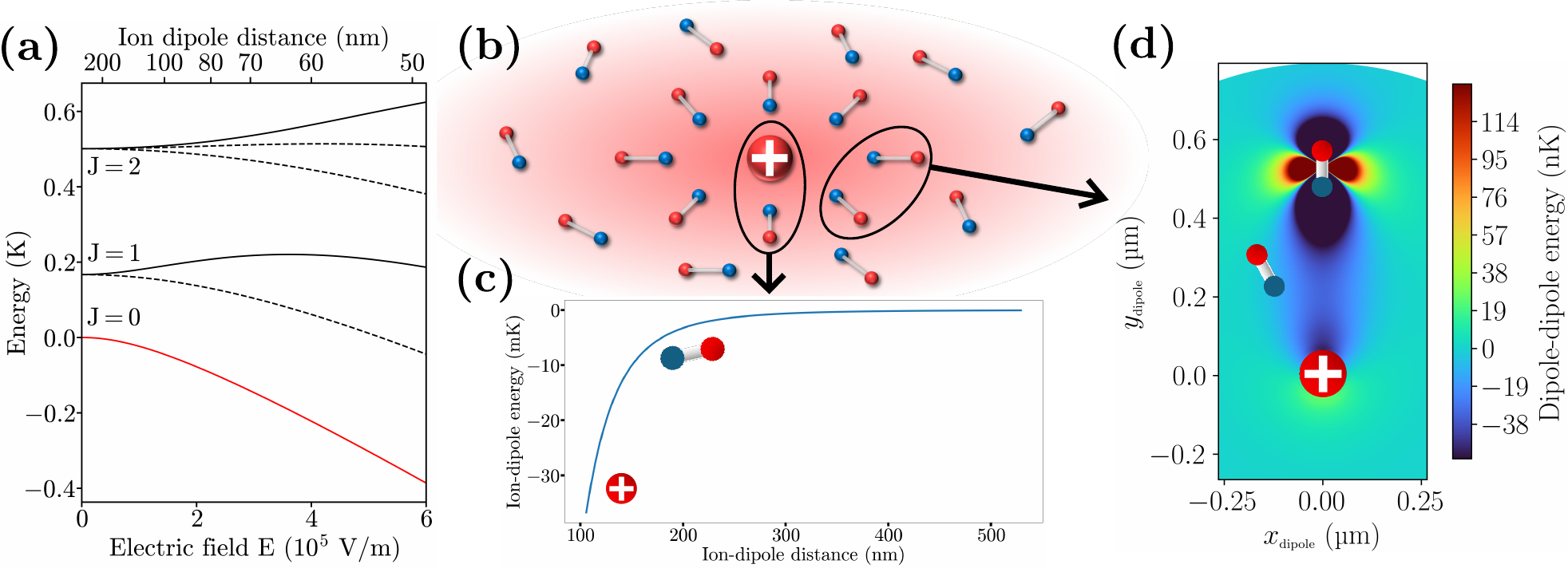}
    \caption{A schematic of our study. Panel (a) shows the Stark manifold for the states $\ket{X^1\Sigma^+,v=0,J=0,1,2,m_j}$ of the molecule in the presence of the ion's electric field. The ground state is marked in red, states where $m_j=0$ are marked with solid lines, and states with non-zero $m_j$ are marked with dashed lines. Panel (b) shows our system: an ion immersed in an 2-dimensional ultracold low-density gas of NaCs molecules. Two interactions are circled: the ion-dipole interaction, and the dipole-dipole interaction. Panel (c) shows the ion-dipole interaction, and (d) shows the anisotropic interaction between two dipoles. At several different field strengths $E$ up to $10^{-6}$ a.u., the ground state energy $U$ (corresponding to $J=0$) is calculated, incorporating rotational states up to $J=10$. For $J=8$ the lowest energy eigenvalue at a field strength corresponding to an ion–dipole distance of 37 nm differs from the $J=10$ value by a very small relative error of $10^{-12}$, and the $J=12$ value differs by an even smaller relative error, so the ion-dipole interaction energy employed in our calculations is converged with respect to $J_\textrm{max}$.} 
    \label{fig:fig1}
\end{figure*}


Using a classically accelerated quantum Monte Carlo method~\cite{TomasReview,Saajid2024}, we calculate the system’s evaporation energy as a function of the number of molecules bound to the ion. This allows us to identify magic numbers corresponding to especially stable configurations. We find that the evaporation energy develops a plateau-like structure, indicating that the molecules arrange in ring-like structures around the ion. These rings stand in stark contrast to those of charged clusters, such as protonated argon or charged van der Waals clusters, which usually follow the dictates of the coordination number of the charged impurity~\cite{Chowdhury2026} or conventional icosahedral packing~\cite{HARRIS1986316}. Here, the structures are governed by the topology of the ion’s electric field and by the anisotropic repulsion between aligned molecular dipoles. Furthermore, we analyze the stability of these structures, including the role of the time-dependent ion trap, and find that many of them should be readily observable under standard temperature and density conditions. From a dynamical perspective, we anticipate that, due to the competition between the attractive ion-molecule force and the repulsive dipole-dipole interaction within a ring, the system is shielded against short-range effects: the dipoles do not reach the short-range region of the ion-molecule interaction potential where chemical reactions occur. Our results therefore establish a new class of mesoscopic molecular ions, stabilized by self-limiting electrostriction, that, in the case of strong ion-dipole interaction, evolve into crystalline molecular rings.

\section{Physical properties of the system}

The system we consider in this work is an ion immersed in an ultracold dilute two-dimensional gas of polar molecules, as schematically represented in Fig.~\ref{fig:fig1}. Each molecule will experience the ion's electric field, perturbing it from its original rovibrational ground state $\ket{v=0,J=0,m_J=0}$, where $v$ and $J$ are the vibrational and rotational quantum numbers, and $m_J$ is the projection of $\vec{J}$ onto the quantization axis, chosen parallel to the E-field produced by the ion. The interaction through the electric field with the dipole moment of the molecule is described by the following Hamiltonian~\cite{stark_effect} (assuming the molecules as rigid rotors and neglecting hyperfine interactions):
\begin{equation}\label{Hamiltonian}
    \hat{H}(r)=B_0\hat{J}^2-d_0E(r)\cos\theta
\end{equation}
where $r$ is the ion-dipole distance, $B_0=\hbar^2/(2I)$ is the rotational constant of the molecule, $\hat{J}^2$ is the rigid rotor dimensionless squared angular momentum operator with eigenvalues $J(J+1)$, $d_0$ is the permanent body-fixed-frame dipole moment of the molecule, $E(r)=\frac{q}{4\pi\epsilon_0 r^2}$ is the ion's electric field, and $\theta$ is the angle between the molecular axis and the electric field. For the NaCs molecule, whose permanent dipole moment is 4.69 Debye and $B_0/h$ is 1.74$\times$10$^9$ Hz, the energies as a function of the electric field of the ion are shown in Fig.~\ref{fig:fig1}(a), where it is observed that molecules in the $J=0$ state behave as high-field seeking states, and hence they will feel an attractive interaction towards the ion, as represented in Fig.~\ref{fig:fig1}(c). It should be noted that if $d_0$ increases, then the Stark manifold compresses horizontally by the same factor, and if $B_0$ increases, then the Stark manifold stretches vertically and horizontally by the same factor; modulo this scaling, the Stark manifold looks the same for any polar molecule in an external electric field. 

In the limit that the ion-molecule distance is large, so their electronic clouds do not overlap, the energy of the molecule interacting with the ion can be written as
\begin{equation}\label{Eigenenergy}
    U(E)=U_{\text{rot}}-d(E)E+\mathcal{O}(E^2).
\end{equation}
Here, $U_\textrm{rot}=B_0J(J+1)$ is what the rotational energy would be in the absence of electric field, $d(E)$ is the molecule's field-dependent effective dipole moment (i.e., the expectation value of the dipole moment projection operator $d_0\cos\theta$ in the lab frame) which we calculate from the Hamiltonian ground state $U$ via
\begin{equation}\label{dipole}
    d(E) = -\frac{\partial U}{\partial E}.
\end{equation}
Hence, the dipole moment of the molecule depends on the electric field of the ion, or, equivalently, on the ion-molecule distance. As a consequence, the bath of polar molecules is described as (in atomic units)
\begin{equation}\label{Vdd}
V_{dd}(\vec{r}_{i},\vec{r}_{j})=\frac{\vec{d}_{i}\cdot\vec{d}_{j}-3(\vec{d}_{i}\cdot\hat{R}_{ij})(\vec{d}_{j}\cdot\hat{R}_{ij})}{|\vec{R}_{ij}|^3},
\end{equation}
where $\vec{d}_{i}=\vec{d}(E(r_i))$, and $\vec{R}_{ij} = \vec{r}_i-\vec{r}_j$ and $\hat{R}_{ij}$ are the vector and unit vector, respectively, representing the distance between the two molecules. It is worth emphasizing that Eq.~(\ref{Vdd}) has the same form as the general dipole-dipole interaction, as depicted in Fig.~\ref{fig:fig1}(d). However, in the present case, the dipole moments depend on the positions of the molecules, a unique feature of polar molecules that is absent in dipolar gases composed of highly magnetic atoms.

\section{Theoretical and computational approach}\label{Theory}

Our approach consists of finding the ground state energy of a system of a single ion interacting with a set of polar molecules. To this end, first we need to draw the general interaction potential $V$ containing one ion and $n$ molecules including ion-dipole and dipole-dipole interactions, given by (in atomic units)
\begin{multline}\label{Total_pot}
    V = \sum_{i=1}^n\Biggl\{ \Biggl[\frac{C_8^{\text{id}}}{|\vec{r}_{\text{ion}}-\vec{r}_i|^8}+U\left(\frac{1}{|\vec{r}_{\text{ion}}-\vec{r}_i|^2}\right) \Biggr] \\
    + \sum_{j>i}^n \Biggl[ \frac{C_8^{\text{dd}}}{|\vec{R}_{ij}|^8} +\frac{\vec{d}_i\cdot\vec{d}_j-3(\vec{d}_i\cdot\hat{R}_{ij})(\vec{d}_j\cdot\hat{R}_{ij})}{|\vec{R}_{ij}|^3} \Biggr]  \Biggr\},
\end{multline}
where $C_8^{\text{id}}$ and $C_8^{\text{dd}}$ are short-range coefficients for the ion-dipole potential and the dipole-dipole potential. This short-range regularization is introduced to represent the unresolved repulsive region and to explore how microscopic length scales influence the shell capacities. At distances where electronic overlap becomes important, the field-dressed long-range ion–molecule and dipole–dipole interactions cease to provide a complete description. As commonly done in theoretical studies of ions immersed in neutral quantum media, we represent the unresolved short-range physics by regularized repulsive cores while retaining the physical long-range interactions explicitly. The $r^{-8}$ form is chosen for computational convenience and is not intended as an ab initio representation of a specific molecular pair. Our principal conclusions do not depend on this particular choice: Table I gives the corresponding ring-capacity relations for general inverse-power and exponential cores, as well as for the limit in which no intermolecular short-range core is present. 




Instead of using the interaction energy at every value of the electric field, it is better to find a functional form for the evolution of the $J=0$ state of the molecule with the strength of the electric field. The functional form that we find optimal to describe the energy of the molecule as a function of the electric field is 
\begin{equation}
    U(E) = \begin{cases}\sum\limits_{k=0}^{5}c_k\left(\frac{E}{E_\textrm{s}}\right)^{2k+2},&E<E_\textrm{t}\\
    U(E_\textrm{t})-d(E_\textrm{t})(E-E_\textrm{t}),&E\geq E_\textrm{t}.\end{cases}
\end{equation}
where $c_k$ ($k=0,1,\ldots,5)$ are fitting parameters, $E_\textrm{t}=0.82\times 10^{-6}$ a.u. is chosen as the threshold electric field for the linear regime, and $E_\textrm{s}=10^{-6}$ a.u. is a fixed electric field scale.



\subsection{Classical optimization}\label{Classical optimization}

Since the gas is ultracold, the molecules will move and orient themselves to the ground state of the Hamiltonian of the full system with potential energy $V$. To find the ground state, we find the global minimum of $V$, using basin hopping (BH) \cite{wales1997}, which does Monte Carlo Metropolis sampling of coordinates according to their nearest local minima of $V$ found by gradient descent. We analytically derive a formula for the gradient of $V$, feeding it to the algorithm to accelerate the gradient descent. 

Keeping the ion fixed at the origin and allowing the dipole positions to vary in the $x,y$ directions (but keeping their orientations pointing towards the origin), we run 14 parallel instances (with different random seeds) of basin hopping for 10000 iterations with BH-temperature 32 nK (not a physical temperature) with position step range $\pm 52.9$ nm. For each basin hopping instance, the initial configuration is $n$ dipoles uniformly randomly spawned in an annulus around the ion with inner radius $r_{\textrm{e},\textrm{id}}$ and outer radius $3r_{\textrm{e},\textrm{id}}$, where $r_{\textrm{e},\textrm{id}}$ is the equilibrium distance of the ion-dipole interaction. 

For the 3D configurations, we use a more advanced global optimization technique which optimizes one cluster at a time, using the optimal configuration for $n-1$ with an additional randomly placed dipole as the initial configuration for $n$, and doing another pass-through taking away the most expensive atom from the $n+1$ configuration as the initial configuration for $n$. A more elaborate version of basin hopping, with adaptive temperature and step-size leading to an average 25\% acceptance rate, and different types of perturbations including angular and radial perturbations, is used for each initial configuration.

The $(x,y)$ positions of the dipoles of the local minimum with the lowest energy are taken to be the global minimum configuration for $V$. These positions are used as the initial configuration for the quantum Monte Carlo approach to calculate the ground state of the system. 

\subsection{Quantum Monte Carlo approach}

Even in the ground state, there is an inherent uncertainty in the positions of the ion and molecules. This leads to a zero-point energy contribution to the quantum mechanical ground-state energy of the system. To calculate the ground-state energy, we use a quantum Monte Carlo technique called diffusion Monte Carlo (DMC), which numerically finds the ground-state solution of the Schrödinger equation by finding the steady-state solution of the \textit{imaginary}-time Schrödinger equation
\begin{equation}
    \hbar\frac{\partial\psi}{\partial\tau}=\sum_{i=0}^n\left(\frac{\hbar^2}{2m_i}\nabla^2\psi\right)-V\psi,
\end{equation}
which is a diffusion equation \cite{kosztin1996}. In this technique, walkers undergo random walks exploring configuration space, promoted via replication at more energetically favorable configurations and penalized via deletion at energetically unfavorable configurations, until their distribution converges to the ground-state wavefunction of the system; the average potential energy of these walkers is then the calculated ground-state energy. 

First, we initialize 2000 walkers in the global minimum configuration given by the classical approach in the previous section. Then, each of 10000 timesteps, all of the walkers perform Gaussian random steps, meaning that for each walker, each Cartesian coordinate of the ion is nudged by a random normal with standard deviation $\sqrt{\hbar d\tau / m_\textrm{ion}}$, and each Cartesian coordinate of each molecule's center of mass is nudged by a random normal with standard deviation $\sqrt{\hbar d\tau / (m_\textrm{Na}+m_\textrm{Cs})}$. Here, $d\tau=5\times 10^{9}$ a.u. is the infinitesimal time increment (on an atomic timescale it is large because it compensates for the small potential energies on the order of $10^{-9}$ $E_h$). Next, comparing the potential energy of the nudged configuration to the current ground state energy estimate, $\textrm{min}(\left\lfloor\exp(-(V-E)d\tau/\hbar)+u\right\rfloor,3)$ replicas of the walker are kept, where $u\in[0,1]$ is uniformly randomly generated. The new energy estimate for this timestep becomes the average potential energy of the new walkers, plus a correction factor to stabilize the walker count \cite{kosztin1996}. Finally, after 10000 timesteps of this process, the average energy estimate of the last 5000 timesteps is reported as the ground state energy. We estimate the standard error of the mean for a single run of DMC using a sample chunking scheme to account for the autocorrelation between adjacent timesteps, and we actually run 14 parallel independent-random-seed instances of DMC and calculate the mean reported energy across the 14 runs.

This quantum Monte Carlo approach gives us insight into the contribution of the zero-point energy of the different particles, allowing us to see the effect of the mass of the ion, for example. Independently of DMC, in the next section we also run a molecular dynamics simulation of the global minimum configuration nudged by initial thermal kicks to analyze the stability of the system.

\subsection{Stability Analysis}\label{section_cd}

Apart from finding the global minimum for a given number of molecules, we investigate the stability of those clusters when the trapping potential is included. Specifically, to simulate the dynamics of the system of a trapped ion in a 2D dipolar gas, we used the DOP853 algorithm~\cite{dop853} with 10$^{-9}$ relative tolerance and 10$^{-5}$ and 10$^{-15}$ absolute tolerance for the position and velocity components, respectively. For the potential, we include both the interaction potential Eq.~\ref{Total_pot} and the trap potentials for both the ion and the molecules, and assume that the effective dipole moments always align with the ion's electric field when computing dipole-dipole forces. For a trapped ion with no molecules around, its equation of motion in the trap is
\begin{equation}
    \ddot{r}_{\text{ion},j} + \frac{\Omega_\textrm{rf}^2}{4}(a_j+2q_j\cos(\Omega_\textrm{rf}t))r_{\text{ion},j}=0,
\end{equation}
where $\Omega_\textrm{rf}=2\pi \times 5$MHz is the RF frequency of the trap, $a_j=\{-3.7,1.8,1.9\}\times10^{-3}$ and $q_j=\{0.283,-0.283,0\}$ are trap parameters determined by the trap voltages and frequency, and the ion’s mass, and the index $j=x,y,z$ labels the Cartesian directions~\cite{Didi}. For the molecules, we also include an optical trapping potential given by
\begin{equation}
    V_{\text{laser}} = -V_0\exp{\left(-\frac{m_d\omega_d^2z^2}{2V_0}\right)},
\end{equation}
where $V_0= k_B \times 1\mu$K is the depth of the trap, $m_d$ is the mass of the molecule, $\omega_d=2\pi\times 50$ kHz is the optical trap frequency, and $z$ is the dipole $z$ coordinate, with the Boltzmann constant $k_B$. Since the trap and interaction potentials are additive, the equations of motion are obtained from their sum.


For the initial state of the system, we first take the lowest-energy geometry we get from the optimization described in Section~\ref{Classical optimization}. The ion's position in the $x$, $y$, and $z$ directions is displaced by random values sampled from Gaussian distributions whose scale is determined by the ion trap's frequency in the corresponding axis. In this paper, we used $ \omega_x=\omega_y=2\pi\times 500$ kHz, and $ \omega_z=2\pi\times 100$ kHz. The dipoles' positions are displaced by random values sampled from a uniform distribution between $-5.29$~nm and $+5.29$~nm. We also give the ion and the dipoles random initial velocities, sampled from Gaussian distributions whose scales are related to the temperature of the ion/dipolar gas, where for the ion we sampled temperatures ranging from 1 $\mu$K to 1 K, and for the dipoles the temperatures are 1 nK. For each trajectory, we simulated it for 1ms and explored different short-range coefficients and different numbers of dipoles to examine the stability of the system, i.e., how easy it is to break the bond between the ion and one dipole in the system.

\section{Results}

\begin{figure}
    \includegraphics[width=0.75\linewidth]{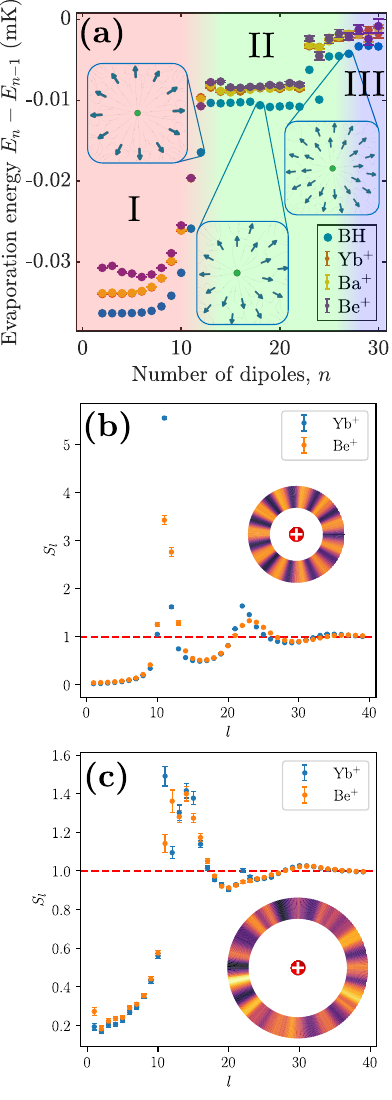}
    \caption{Evaporation energies and angular structure factors $S_l$ for 2D NaCs clusters surrounding an ion. (a) Basin-hopping (BH) and DMC evaporation energies; the insets show the first magic number, an intermediate cluster, and the second magic number. (b,c) Angular structure factors and DMC angular distributions for the inner and outer rings of the second-magic-number cluster, respectively. Here, $C_{8,\mathrm{id}}=10^{22}$ a.u., $C_{8,\mathrm{dd}}=5\times10^{18}$ a.u., $r_{e,\mathrm{id}}=520$ nm, and $d_{\mathrm{dd}}=270$ nm.}
    \label{fig:fig2}
\end{figure}

The main result of this work is presented in Fig.~\ref{fig:fig2}, displaying the evaporation energy as a function of the number of molecules for three different ions. Throughout our results, we report energies in units of temperature; for example, evaporation energies reported in mK are actually measured in $k_B\times\textrm{mK}$. First, we notice that the evaporation energy exhibits a plateau-like behavior as a function of the number of molecules, in stark contrast to other charged clusters, such as protonated argon clusters~\cite{Chowdhury2026} or ions in H$_2$ clusters~\cite{H2clusters}. The ion’s electric field aligns and attracts the polar molecules, producing an electrostrictive increase in molecular density around the impurity. As the first ring fills, the growing intra-ring repulsion limits further densification and eventually makes occupation of an outer ring energetically favorable. We refer to this repulsion-limited aggregation as self-limiting electrostriction. This mechanism gives rise to the formation of mesoscopic molecular ions where a single ion binds several molecules.

Because of the plateau structure, the conventional identification of magic numbers from local maxima in the evaporation energy is no longer appropriate. Since each plateau corresponds to the population of a new molecular ring, we instead define a magic number as the cluster size at which a ring is completed such that the next molecule preferentially occupies a second ring. Representative configurations are shown in insets I and III of Fig.~\ref{fig:fig2}, while inset II corresponds to an intermediate cluster size between two successive ring closures. Interestingly, the DMC calculations reveal exchange of molecules between neighboring rings, producing quantum delocalization over several accessible structural arrangements. Consequently, the ring-closure magic numbers inferred from the DMC density distributions need not coincide exactly with those predicted by the classical BH configurations. 

In addition, a noticeable difference exists between the basin-hopping (BH) and DMC evaporation energies, regardless of the cluster size. This difference demonstrates the importance of zero-point motion and quantum delocalization in determining the energetics of these clusters. Indeed, the mass of the ion plays a role in the first ring, as the large variation of the evaporation energy shows, while its effect washes out upon formation of the second ring. The reason for this behavior is that, once the first ring is completed, the structure becomes more rigid and the zero-point energy of the ion is suppressed. Equivalently, the dipolar ring acts as a trapping potential for the ion.

The plateau-like structures on the evaporation energy as a function of the number of molecules indicate an underlying binding mechanism. To explore this further, we compute the angular structure factor given by
\begin{equation}
\label{Sl}
S_l=\frac{1}{N_d}\langle |\sum_{i=1}^{N_d}e^{\imath l \theta_i}|^2\rangle,
\end{equation}
which can be interpreted as the Fourier transform of the angular distribution function of the particles. In Eq.~(\ref{Sl}) $\theta_j$ represents the polar angle of the j-dipole within a given ring containing $N_d$ molecules and $l$ is an integer corresponding to angular Fourier mode.

The angular structure factors associated with the first and second rings are shown in panels (b) and (c) of Fig.~\ref{fig:fig2}, for two different ionic species. Panel (b) displays the angular structure factor for the first ring, which exhibits a strong peak at $l=11$, corresponding to the most frequently observed ring size instead of the BH magic number of 12. However, the peak for Yb$^+$ is more pronounced than that for Be$^+$, indicating that the former produces a more rigid and crystal-like structure than the latter. In both cases, several additional Fourier modes acquire nonzero amplitudes, indicating structural flexibility. This behavior can also be inferred from the inset, which shows the angular distribution function of the dipoles in the first ring. The second ring, shown in the lower panel, exhibits markedly different behavior. Although a peak remains near the corresponding magic number, it is broader and weaker than the peak associated with the first ring. More importantly, the results are nearly independent of the ionic species, in agreement with the evaporation-energy results presented in Fig.~\ref{fig:fig2} and consistent with a more flexible structure. The second ring is therefore more flexible than the first, primarily because the induced dipole moment weakens as the molecules move farther away from the ion. Along the same lines, we attribute the insensitivity to the ionic species to the reduced rigidity of the second ring. More importantly, these results confirm that the characteristic dependence of the evaporation energy on the number of dipoles is a rather general feature of two-dimensional ion–dipole many-body systems.

\begin{figure*}
    \includegraphics[width=\linewidth]{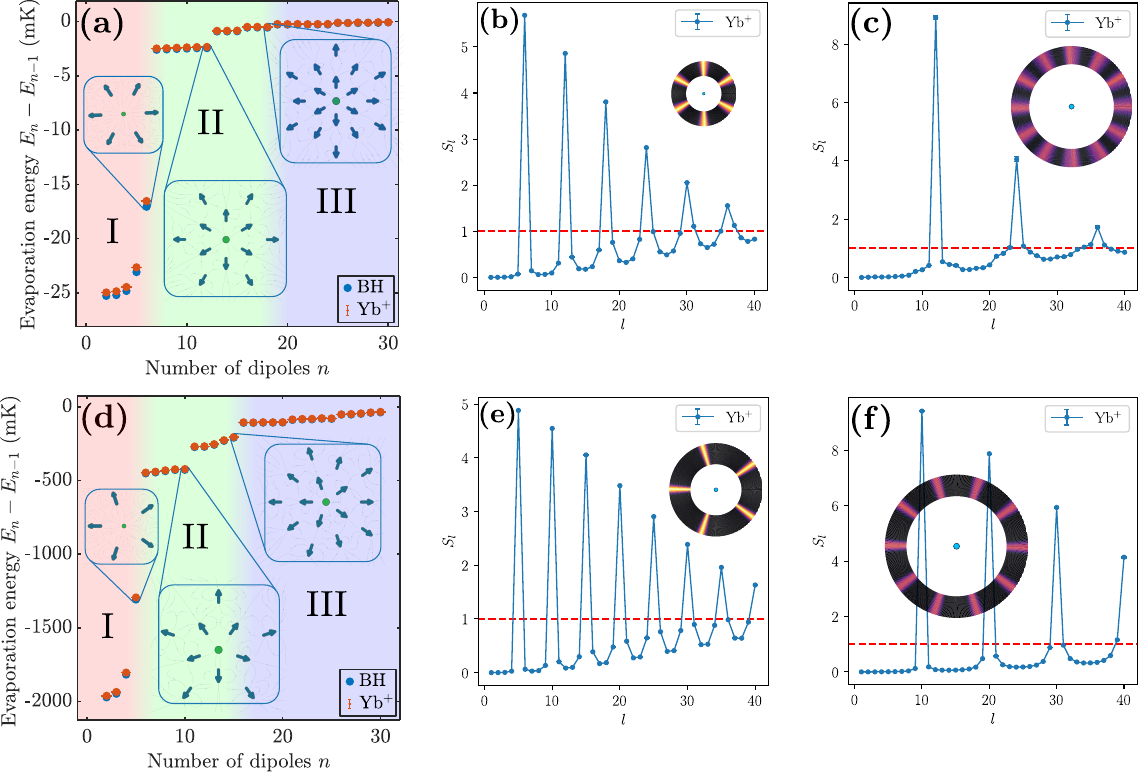}
    \caption{Evaporation energies and angular structure factors $S_l$ for 2D NaCs clusters surrounding an ion. ``BH'' denotes basin-hopping global minima, while ``Yb$^+$'' denotes DMC results for a Yb$^+$ ion. Panels (a,d) show the evaporation energies, with insets illustrating the first magic number, an intermediate cluster, and the second magic number. Panels (b,e) and (c,f) show the angular structure factors and DMC angular distributions of the inner and outer rings of the second-magic-number cluster, respectively. Panels (a--c) correspond to $C_{8,\mathrm{id}}=10^{19}$ a.u., $C_{8,\mathrm{dd}}=10^{18}$ a.u., $r_{e,\mathrm{id}}=100$ nm, and $d_{\mathrm{dd}}=102$ nm; panels (d--f) correspond to $C_{8,\mathrm{id}}=C_{8,\mathrm{dd}}=10^{15}$ a.u., $r_{e,\mathrm{id}}=20$ nm, and $d_{\mathrm{dd}}=24$ nm.}
    \label{fig:fig3}
\end{figure*}

It is necessary to understand if the plateau-like behavior of the evaporation energy is a universal phenomenon. To this end, we have computed the evaporation energies versus the number of molecules of two weaker ion-dipole and dipole-dipole short-range potentials, and the results are shown in Fig.~\ref{fig:fig3}. The figure clearly shows that the plateaus are a general trend of the system under consideration, and the different regions highlighted in Fig.~\ref{fig:fig2} survive. As $C_8^\textrm{id}$ decreases, the ion–molecule equilibrium radius becomes smaller, reducing the circumference of the first ring and therefore its molecular capacity. 

The angular structure factors for the first ring [panels (b) and (e)] and second ring [panels (c) and (f)] are shown in Fig.~\ref{fig:fig3}. The first-ring structure factor exhibits pronounced peaks at integer multiples of the fivefold angular order, which indicates a very rigid and crystal-like structure. The same can be seen from the angular distribution, showing a narrow angular displacement of the dipoles within the ring. The second ring is slightly less rigid and the dipoles show some fluctuation around their equilibrium. The rigidity of the structures can be explained in light of the self-limiting electrostriction. As the ion-dipole interaction shows a weaker short-range repulsion, the equilibrium distance is smaller and the depth of the interaction potential deeper. This stronger attraction brings the molecules closer to the ion, where they acquire a larger effective dipole moment, and, with it a stronger repulsion that ultimately translates into a stronger angular localization.

\subsection{The role of short-range physics} 
\begin{figure}
    \includegraphics[width=0.9\linewidth]{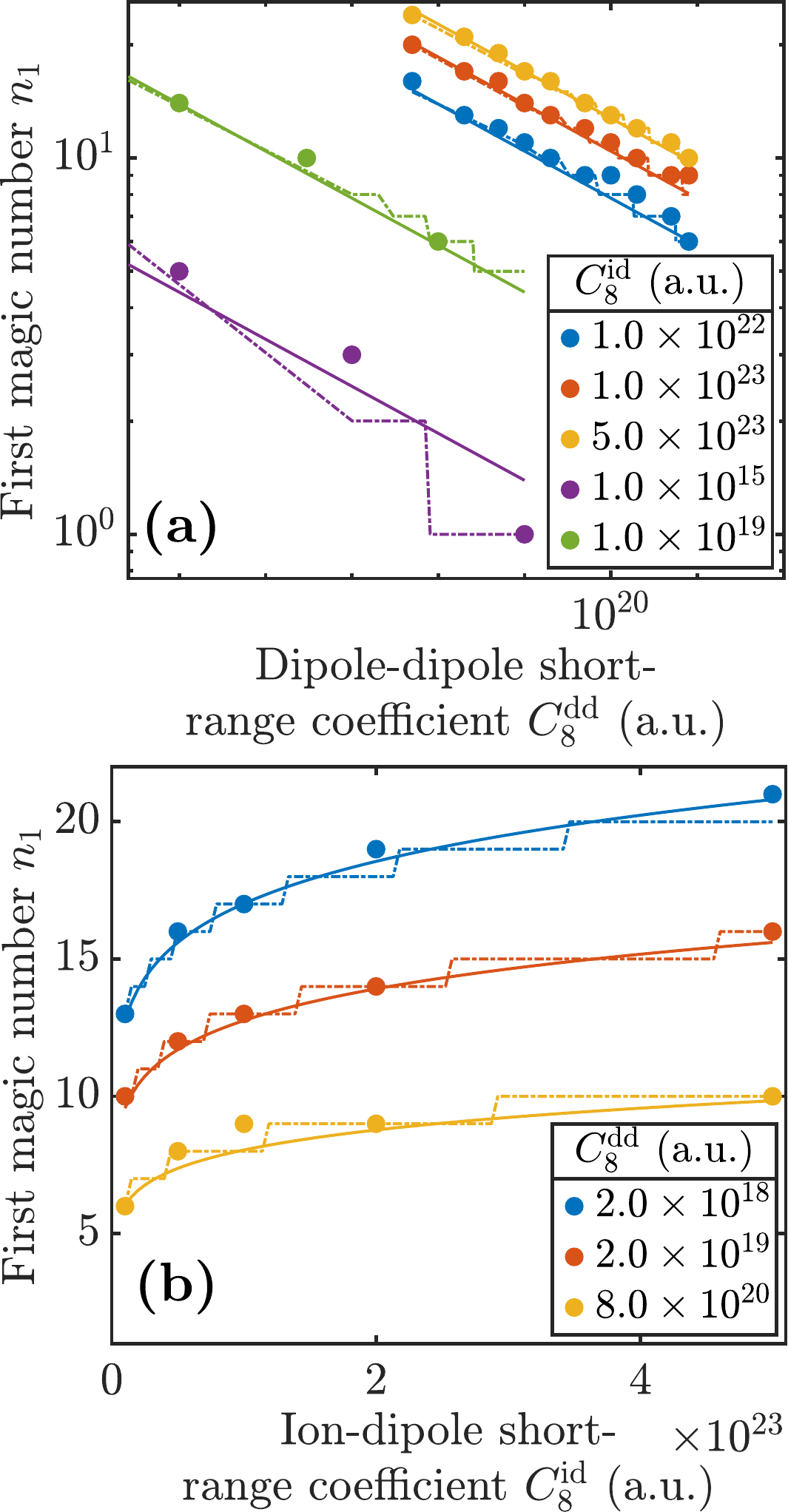}
    \caption{Magic numbers of dipoles within the first ring surrounding the ion, as a function of the short-range interaction coefficients $C_8^\textrm{dd}$ and $C_8^\textrm{id}$, computed by basin-hopping. The dotted lines show accurate predictions for the magic number according to Eqs.~\ref{eq:Vring},\ref{eq:inequality}, sweeping one $C_8$ parameter for three different values of the other $C_8$ parameter. The solid curves show approximate asymptotic predictions given by Eq.~\ref{eq:n}.}
    \label{fig:fig4}
\end{figure}

Given the accuracy of BH in describing the magic numbers of the system at hand, we now turn to the specific structure of each ring and the effects of short-range physics. We computed the number of molecules in the first dipolar ring as a function of the short-range coefficients for ion-dipole and dipole-dipole interaction potentials, and the results are shown in Fig.~\ref{fig:fig4}. Increasing $C_8^\textrm{id}$ moves the ion–molecule equilibrium distance outward and enlarges the first ring, whereas decreasing $C_8^\textrm{dd}$ reduces the intermolecular exclusion distance. Both changes allow more molecules to occupy the first ring. The equilibrium distance for the ion-dipole interaction potential sets the radius of the first ring, but for dipole-dipole interactions it is the short-range repulsion that establishes the intermolecular distance within the ring, showing a more important contribution to the magic number. Therefore, short-range interactions affect the magic numbers, but they do not affect the overall plateau-like behavior of the evaporation energy as a function of the number of molecules.

\subsection{A model for the magic numbers}

Assume that the first magic number geometry is a circular ring of $n$ dipoles surrounding the ion at radius $r_e$, the equilibrium distance of the ion-dipole interaction. The $i^\textrm{th}$ dipole in this ring is located at $\vec{r}_i=(r_e\cos\frac{2\pi i}{n}, r_e\sin\frac{2\pi i}{n}, 0)$. Substituting this into Eq.~\ref{Total_pot}, we obtain the potential energy of the ring as
\begin{multline}\label{eq:Vring}
    V_\textrm{ring}(n)\approx -nD_e\\
    +n^9\frac{C_8^\textrm{dd}}{r_e^8}\frac{1}{512\cdot 14175}\Bigg(3+\frac{40}{n^2}+\frac{294}{n^4}+\frac{2160}{n^6}-\frac{2497}{n^8}\Bigg)\\
    +n^4\frac{d(r_e)^2}{r_e^3}\frac{1}{16}\Bigg(\frac{2\zeta(3)}{\pi^3}+\frac{3}{\pi n^2}\left(\log\frac{2n}{\pi}+\gamma\right)\\
    -\frac{1}{6\pi n^2}-\frac{37\pi}{720n^4}\Bigg),
\end{multline}
 The first row is just $n$ times the ion-dipole dissociation energy $D_e$, since every dipole is at distance $r_{e,id}$ from the ion. The second row is the sum of the dipole-dipole short-range repulsive interaction, which scales as $\mathcal{O}(n^9)$ since $n$ dipoles on a ring have nearest-neighbor distance approximately $2\pi r_e/n$, which means that the nearest-neighbor dipole-dipole terms $C_8/r^8$ scale as $n^8$, and there are $n$ such nearest-neighbor pairs. The third row is the sum of the dipole-dipole interaction, which scales as $\mathcal{O}(n^4)$ since the $n$ nearest-neighbor pairs of dipoles interact with energy $d^2/r^3$ where $r\approx 2\pi r_e/n$. Note that the leading terms in the series are consistent with the nearest-neighbor distance being roughly $2\pi r_e/n$, since $3/(512\cdot14175)\approx 1/(2\pi)^8$ and $2\zeta(3)/(16\pi^3)\approx 1/(2\pi)^3$.

The first magic number is the smallest number $n$ such that it becomes energetically more favorable to put a dipole outside of the ring rather than to make room for it within the ring:
\begin{equation}\label{eq:inequality}
    V_\textrm{ring}(n)+E_\textrm{new}(n)<V_\textrm{ring}(n+1),
\end{equation}
where $E_\textrm{new}(n)$ is the energy of having a molecule outside the ring, as explained in the Appendix. For several different combinations of $C_8^\textrm{dd}$ and $C_8^\textrm{id}$, we solve for the smallest value of $n$ satisfying this inequality, giving us predictions for the first magic number plotted as dotted lines in Fig.~\ref{fig:fig4}. This ring model, assuming fixed $r_e$ throughout the assembly of the ring and assuming no distortion when a new dipole joins just outside the ring, is able to reproduce most of the magic numbers calculated with BH.

At an equilibrium distance of $r_e$, the depth of the ion-dipole potential is $D_e$. Also, the sum of the dipole-dipole interactions becomes dominated by the sum of the $n$ nearest-neighbor dipole pairs' $C_8^\textrm{dd}$ repulsion. The nearest-neighbor distance is approximately $2\pi r_e/n$, so the sum of these dipole-dipole interactions is approximately $nC_8^\textrm{dd}/(2\pi r_e/n)^8=n^9C_8^\textrm{dd}/r_e^8\cdot1/(2\pi)^8$, which is consistent with the leading term in the second row in Eq.~\ref{eq:Vring}. The non-nearest-neighbor dipoles have a much smaller repulsion since $1/r^8$ falls off quickly. Furthermore, for all $C_8$ combinations that we considered, we find that this $C_8^\textrm{dd}$ repulsion is at least one order of magnitude larger than the sum of the $n$ nearest-neighbor-pair dipolar repulsions, which is $n d(r_e)^2/(2\pi r_e/n)^3=n^4d(r_e)^2/r_e^3\cdot 1/(2\pi)^3$, consistent with the leading term in the third row of Eq.~\ref{eq:Vring}. Thus, $V_\textrm{ring}(n)$ can be written more concisely as
\begin{equation}
    V_\textrm{ring}(n)\approx -nD_e+n^9\frac{C_8^\textrm{dd}}{(2\pi r_e)^8}
\end{equation}
An additional $(n+1)^\textrm{th}$ dipole will energetically prefer to be outside of the first ring instead of inserting itself into the ring when the energy cost to insert a new dipole into the ring, $9n^8C_8^\textrm{dd}/(2\pi r_e)^8$, becomes comparable to the ion-dipole energy $D_e=C_4^2/(4C_8^\textrm{id})$. Assuming that the insertion energy cost equals $0.5D_e$ and substituting $r_e=\sqrt[4]{2C_8^\textrm{id}/C_4}$, we find that $C_4$ cancels out and 
\begin{equation}\label{eq:n}
    n\approx 4.4\left(\frac{C_8^\textrm{id}}{C_8^\textrm{dd}}\right)^{1/8}.
\end{equation}
The predictions for the first magic number according to Eq.~\ref{eq:n} are shown in Fig.~\ref{fig:fig4}, and they agree to within $\pm 1$ for all cases we considered. We have generalized Eq.~(\ref{eq:n}) to accommodate other short-range interaction potentials, and the results are shown in Table~\ref{table}. 

Importantly, the formation of concentric molecular rings does not require the phenomenological intermolecular short-range repulsion. When $C_8^{dd}=0$, the attractive ion–molecule interaction confines the molecules near the equilibrium radius, while the repulsive intra-ring dipole–dipole interaction distributes them angularly around the ion and limits the ring capacity. In this limit, the first magic number is estimated as
\begin{equation}
    n\approx \pi\left(\frac{D_er_e^3}{d(r_e)^2}\right)^{1/3}.
\end{equation}
Thus, the short-range coefficient $C_8^{dd}$ modifies the shell capacity but is not required for the underlying ring-forming mechanism. These results are summarized in Table~\ref{table} that presents a set of analytical expressions for the prediction of the first magic number with its given range of applicability.


\begin{table*}[ht]
    \caption{\label{table}%
    Magic-number formulas for three short-range repulsion models,
    with their domains of validity. The numerical constants in front of the expressions in the first three rows, and the 2 inside the argument of $W_{-1}$ in the last row, are actually $\mathcal{O}(1)$ free parameters which come from the assumption that the cost of adding one more dipole to the ring is 50\% of the ion-dipole well depth.}
    \renewcommand{\arraystretch}{1.4}  
    \setcellgapes{6pt}                 
    \makegapedcells
    \begin{ruledtabular}
    \begin{tabular}{ccc}
        Potentials & Magic number & Validity\\
        \hline
        \makecell{$V_\textrm{id}(r)=\dfrac{C_8^\textrm{id}}{r^8}+U(r)\approx
            \dfrac{C_8^\textrm{id}}{r^8}-\dfrac{C_4}{r^4}$\\
        $V_\textrm{dd}(r)=\dfrac{C_8^\textrm{dd}}{r^8}
            +\dfrac{\vec{d}_i\cdot\vec{d}_j
            -3(\vec{d}_i\cdot\hat{r})(\vec{d}_j\cdot\hat{r})}{r^3}$} &
        $n\approx 4.4\left(\dfrac{C_8^\textrm{id}}{C_8^\textrm{dd}}\right)^{1/8}$ &
        $\dfrac{C_8^\textrm{dd}}{(2\pi r_e/n)^8}\gg
            \dfrac{d(r_e)^2}{(2\pi r_e/n)^3}$\\
        \hline
        \makecell{$\min\limits_r V_\textrm{id}(r)=V_\textrm{id}(r_e)=-D_e$\\
        $V_\textrm{dd}(r)=
            \dfrac{\vec{d}_i\cdot\vec{d}_j
            -3(\vec{d}_i\cdot\hat{r})(\vec{d}_j\cdot\hat{r})}{r^3}$} &
        $n\approx \pi\left(\dfrac{D_er_e^3}{d(r_e)^2}\right)^{1/3}$ &
        $V_\textrm{dd}(2\pi r_e/n)\approx
            \dfrac{d(r_e)^2}{(2\pi r_e/n)^3}$\\
        \hline
        \makecell{$\min\limits_r V_\textrm{id}(r)=V_\textrm{id}(r_e)=-D_e$\\
        $V_\textrm{dd}(r)=\dfrac{C_a^\textrm{dd}}{r^a}
            +\dfrac{\vec{d}_i\cdot\vec{d}_j
            -3(\vec{d}_i\cdot\hat{r})(\vec{d}_j\cdot\hat{r})}{r^3}$} &
        $n\approx\dfrac{2\pi}{\sqrt[a]{2(a+1)}}
            \left(\dfrac{D_e r_e^a}{C_a^\textrm{dd}}\right)^{1/a}$ &
        $\dfrac{C_a^\textrm{dd}}{(2\pi r_e/n)^a}\gg
            \dfrac{d(r_e)^2}{(2\pi r_e/n)^3}$\\
        \hline
        \makecell{$\min\limits_r V_\textrm{id}(r)=V_\textrm{id}(r_e)=-D_e$\\
        $V_\textrm{dd}(r)=Ae^{-Br}
            +\dfrac{\vec{d}_i\cdot\vec{d}_j
            -3(\vec{d}_i\cdot\hat{r})(\vec{d}_j\cdot\hat{r})}{r^3}$} &
        $n\approx\dfrac{2\pi Br_e}
            {\left|1+W_{-1}\!\left(-\dfrac{D_e}{2eA}\right)\right|}$ &
        \makecell{$Ae^{-2\pi Br_e/n}\gg
            \dfrac{d(r_e)^2}{(2\pi r_e/n)^3}$\\[2pt]
        $A>D_e/2$}\\
    \end{tabular}
    \end{ruledtabular}
\end{table*}


\subsection{Larger clusters}

\begin{figure}
    \includegraphics[width=\linewidth]{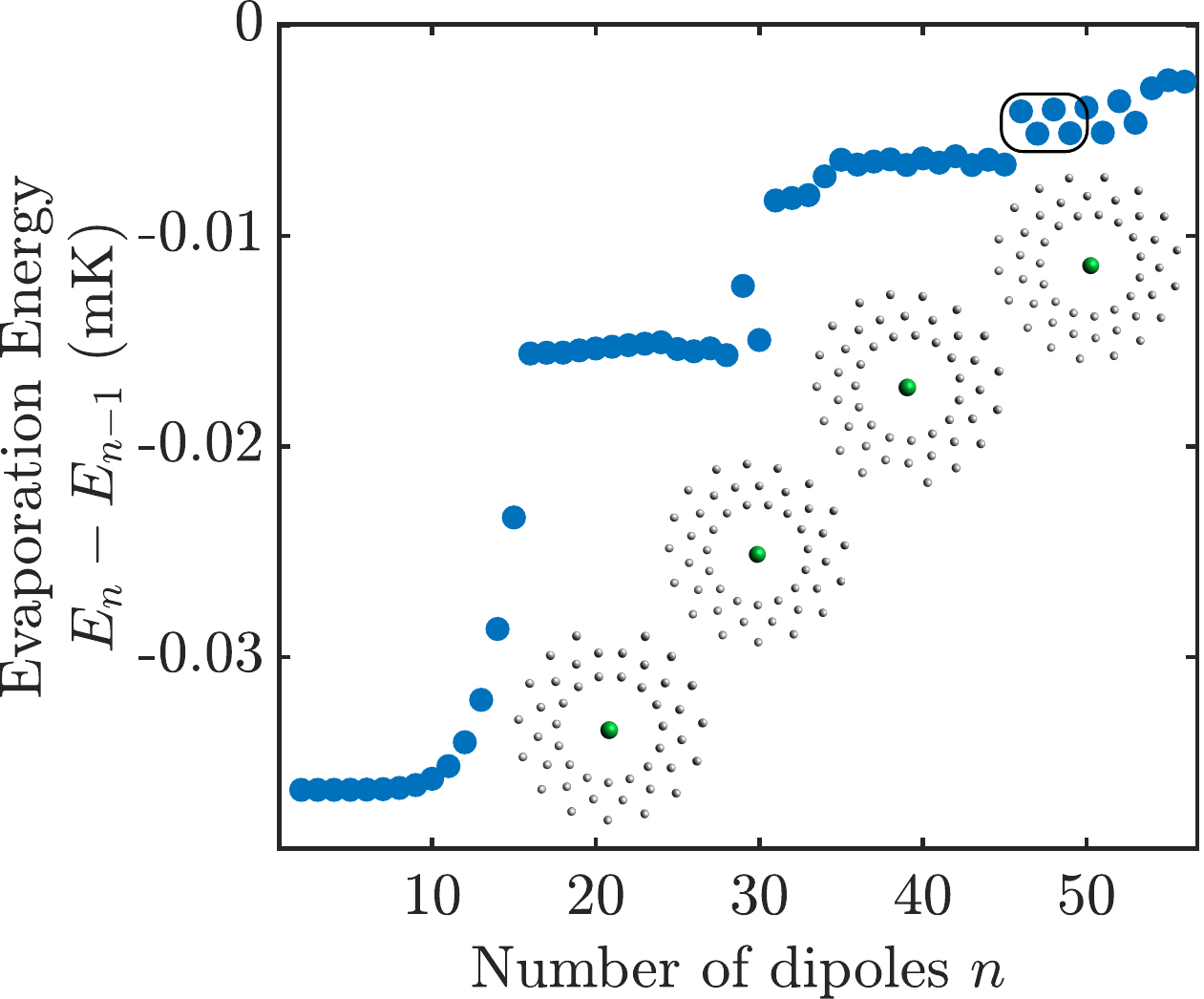}
    \caption{Evaporation energies of 2D clusters of NaCs dipolar molecules surrounding an ion for $n\leq 56$ computed by basin hopping. The ion-dipole equilibrium distance is $9800a_0$, and the dipole-dipole nearest-neighbor distance within the first ring is $4000a_0$. The insets show geometries for $n=46,47,48,49$.}
    \label{fig:fig5}
\end{figure}

For almost every combination of $C_8$ coefficients and larger clusters, we found that, at the end of constructing the third ring of dipoles, the evaporation energy, within the BH approach, oscillates with respect to $n$, as shown in Fig.~\ref{fig:fig5}. In addition, for larger clusters the plateau-like structure in the evaporation energy is smeared out, yielding a more monotonic curve. This is because dipoles are added to the second ring, then the third, then the second again, then the third again, and so on, pointing toward a more disordered structure than in the previous two rings. These dipoles are not immediately added to the outside to build a fourth ring because the ion-dipole attraction is weaker there. It becomes energetically favorable to add a dipole to the second ring, whose intra-ring repulsion is compensated by increasing the circumference of the third ring, thereby relaxing the third ring's intra-ring repulsion. Then, since the third ring's density has decreased (or, since a slot opens up on the third ring, as shown in two of the clusters in Fig.~\ref{fig:fig5}), it becomes favorable to insert a dipole into the third ring, and the second ring once again becomes the least dense, favoring a new dipole to enter the second ring. This alternating-evaporation-energy effect is a hallmark of the competition between ion-dipole attraction and dipole-dipole repulsion. 


\subsection{The onset of mesoscopic molecular ions}

For the calculation of the interaction potential, Eq.~(\ref{Total_pot}), we neglect the back-action arising from dipole–dipole interactions. Specifically, we assume that, in all scenarios considered, the electric field experienced by a given dipole is dominated by the ionic charge rather than by neighboring molecular dipoles. To evaluate the validity of determining each molecular dipole moment solely from the ionic field, we compare the electric field generated by the surrounding molecules—computed using $d(r_e)$, with the field produced by the ion, introducing
\begin{equation}
\label{etaN}
\eta_n=\frac{|\vec{E}_{n\text{dip}}(n,r_i)|}{|\vec{E}_{\text{ion}}(r_i)|}.
\end{equation}
This establishes the ratio between the magnitude of the electric field created by surrounding dipoles and the electric field of the ion. The ratio depends on the number of molecules $n$, and for simplicity we will consider only molecules placed in a ring whose radius is the equilibrium distance of the ion-molecule potential, as in our derivation of the magic numbers. When $\eta_n\ll 1$, replacing the self-consistent local field with the ion field changes the field-dressed molecular dipole only perturbatively.

\begin{figure}
    \includegraphics[width=\linewidth]{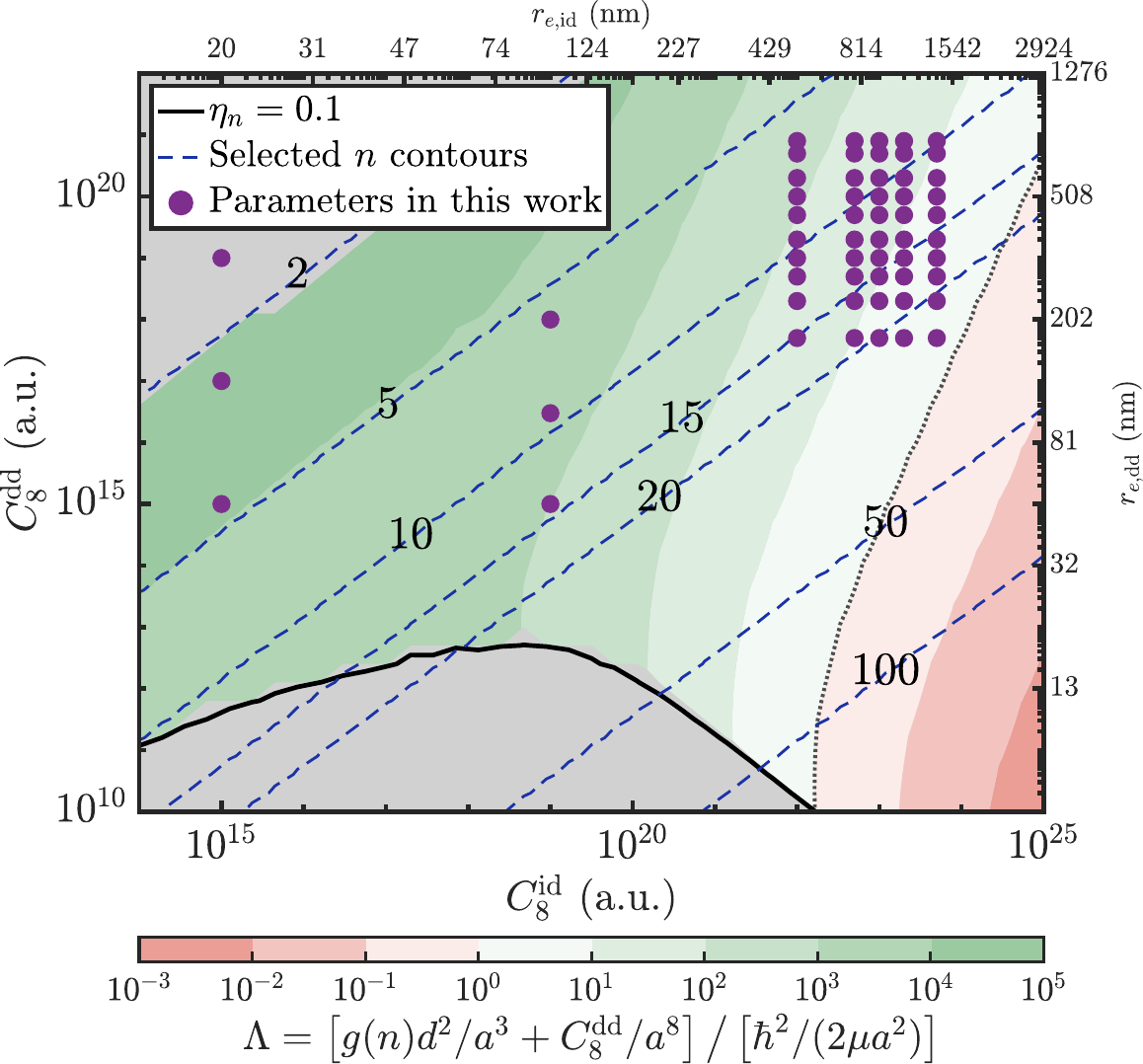}
    \caption{Parameter range of validity of our approach. The points show the values of $C_8^\textrm{id}$ and $C_8^\textrm{dd}$ that we used in our work. The colored area represents the range of $C_8$ values for which the electric field experienced by a molecule in the first ring due to other molecules within the ring is less than 10\% of the electric field experienced due to the ion; i.e., $\eta<0.1$. The gray area at the bottom shows where $\eta>0.1$. The dotted contours show the predictions for the magic numbers given by Eq.~\ref{eq:n}, valid in the colored region, which are used to calculate $\eta$. The colored region indicates how strong the nearest-neighbor dipole-dipole interaction is compared to the azimuthal kinetic energy scale ($a=2\pi r_e/n$, where $r_e$ is the ion-dipole equilibrium distance and $n$ is the prediction for the first magic number. $g(n)=1+\sin^2(\pi/n)$ is an angular factor that accounts for the relative orientation of the nearest-neighbor dipoles).}
    \label{fig:fig6}
\end{figure}

The behavior of $\eta_n$ as a function of the short-range coefficients is shown in Fig.~\ref{fig:fig6}. We can clearly distinguish two regions: a colored region, which denotes the set of short-range parameters for which $\eta_n\le 0.1$—the threshold we deem acceptable—and a gray region, where our approximation fails. The available region spans several orders of magnitude in the short-range coefficients.

To analyze the nature of the ion-molecule structures, we introduce the ratio of the nearest-neighbor dipole-dipole interaction to the azimuthal kinetic energy scale given by 
\begin{equation}
\label{eq_Lambda}
\Lambda=\frac{V_{dd}}{E_{kin}}=\frac{g(n)d^2/a^3+C_8^{dd}/a^8}{\frac{\hbar^2}{2\mu a^2}},
\end{equation}
as the figure of merit. In Eq.~(\ref{eq_Lambda}), $g(n)=1+\sin^2(\pi/n)$ is an angular factor that accounts for the relative orientation of the nearest-neighbor dipoles, $a=2\pi r_e/n$, $r_e$ is the ion-dipole equilibrium distance, $n$ is the predicted first magic number, $d$ is the dipole moment of the molecule evaluated at the electric field of the ion at a distance $r_e$, and $\mu$ is the two-body reduced mass. $\Lambda$ measures the angular rigidity of the structure, and when $\Lambda\gg1$ we can call it a crystalline molecular ring. Therefore, the red areas of Fig.~\ref{fig:fig6} denote flexible structures associated with mesoscopic molecular ions, whereas the dark green areas correspond to crystalline molecular rings.

Therefore, our assumptions hold over a wide range of conditions: from weakly interacting ion-dipole and flexible structures to the strongly interacting regime of crystalline molecular rings. On the right-hand side of Fig.~\ref{fig:fig6}, we are in the weakly interacting regime, where dipoles bound to the ion form mesoscopic ions of size $\gtrsim$500~nm. As we move to the left, we enter an intermediate regime in which dipole-dipole interactions frustrate the ion's pull, inducing more rigid structures. Finally, on the left-hand side of the figure, we enter the strongly interacting regime, where molecules move little around their equilibrium positions, and the competition between ion attraction and dipolar repulsion gives rise to crystalline molecular rings.



\subsection{Stability of ion-dipole clusters}

\begin{figure*}[t]
   \centering
   \includegraphics[width=\textwidth]{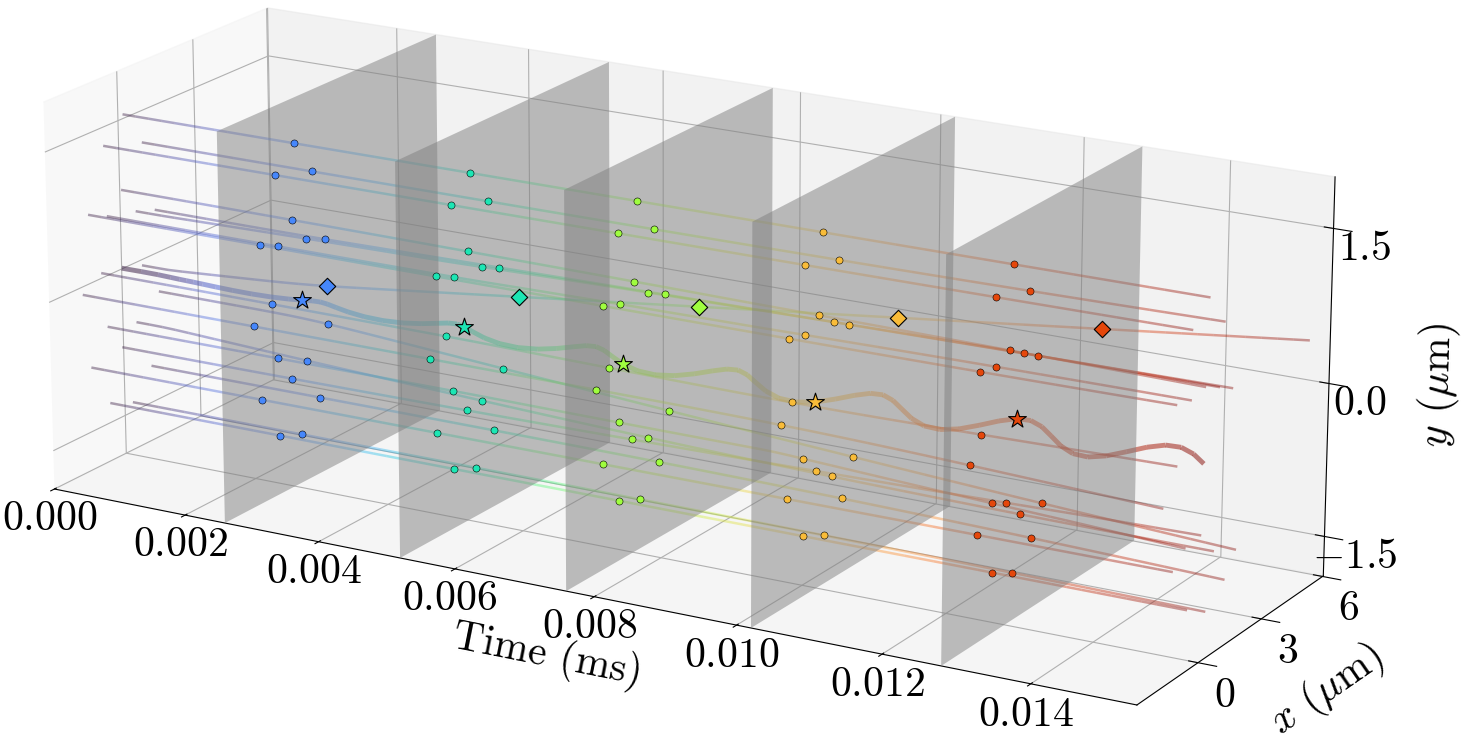}
   \caption{Projection onto the $xy$ plane of an unstable trajectory as a function of time for a cluster of size 19, with $C_8^{\text{id}} = 5 \times 10^{22}$ and $C_8^{\text{dd}} = 8 \times 10^{20}$ in atomic units. Thin curves trace the trajectories of the dipoles, while the thick curve traces the ion trajectory. The cluster configurations at selected times are shown on the gray planes, where circles denote dipoles, the diamond denotes the dipole that dissociates from the cluster, and the star denotes the ion. The color scale indicates the progression of time along the trajectory.} 
   \label{fig:traj}
\end{figure*}

In this section, we examine the stability of trapped ion–dipole clusters against thermal perturbations. We consider three combinations of the short-range interaction parameters and 12 initial ion temperatures. For each combination of interaction parameters and ion temperature, we propagate 96 independent trajectories. The initial positions and velocities of the ion and dipoles are randomly sampled according to their respective temperatures and the trap parameters, as described in Sec.~\ref{section_cd}. A trajectory is classified as unstable if at least one dipole reaches a distance of $10^{5}a_0$ from the ion before the end of the simulation at 1~ms. Otherwise, the trajectory is classified as stable.

A representative dissociation trajectory is shown in Fig.~\ref{fig:traj}, which illustrates the time evolution of the cluster projected onto the $xy$ plane. Initially, the ion oscillates within the cluster, while the dipoles remain localized near their equilibrium positions. As the ion approaches one of the dipoles, marked by a diamond, the ion–dipole separation decreases until the short-range repulsive interaction becomes dominant. The resulting momentum transfer provides the dipole with sufficient kinetic energy to overcome the evaporation threshold and escape from the cluster. Once ejected, the dipole rapidly moves away from the ion, leading to irreversible dissociation of the cluster.

\begin{figure*}[t]
   \centering
   \subfloat[]{\includegraphics[width=0.33\textwidth]{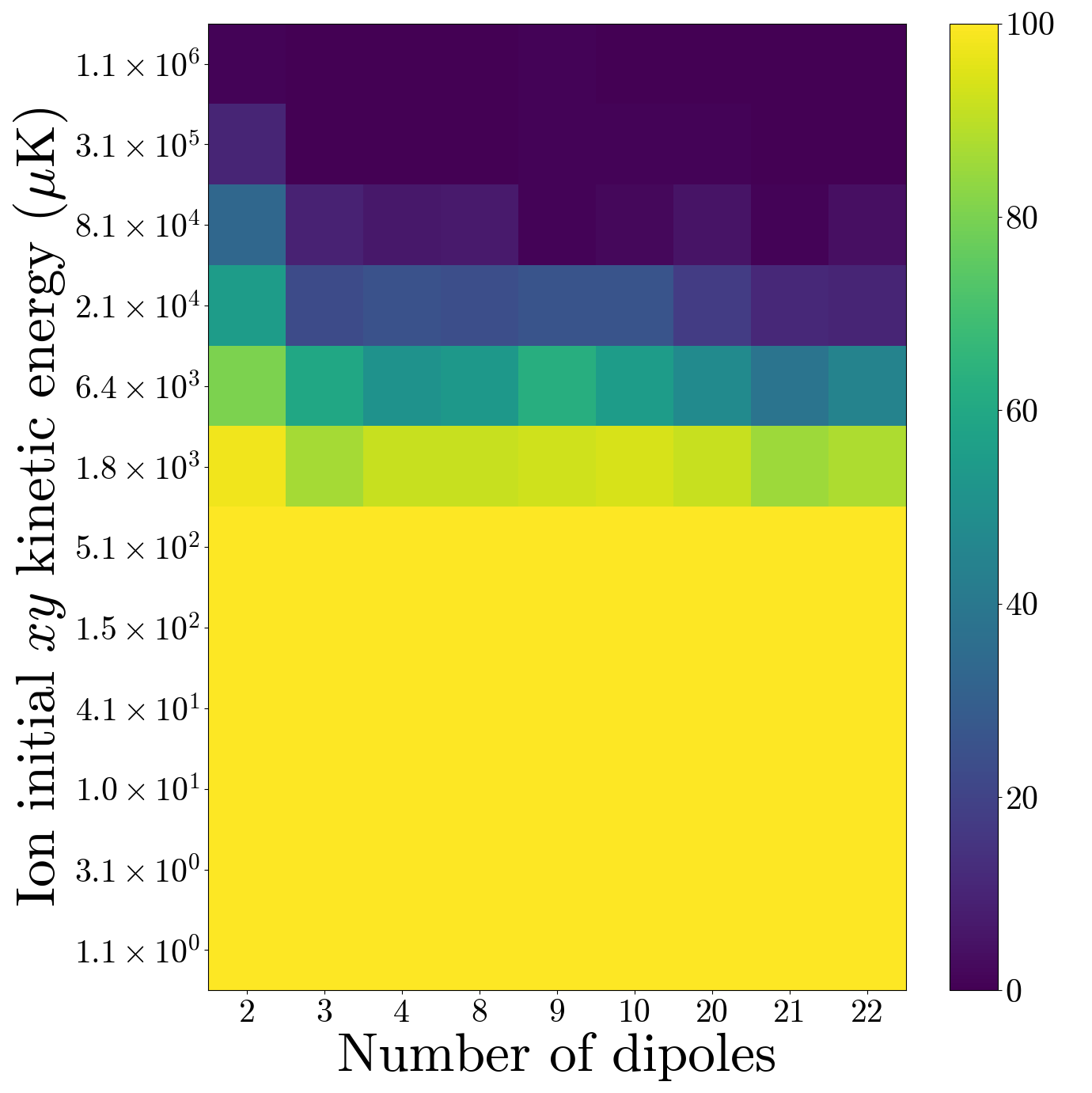}}
   \hfill
   \subfloat[]{\includegraphics[width=0.33\textwidth]{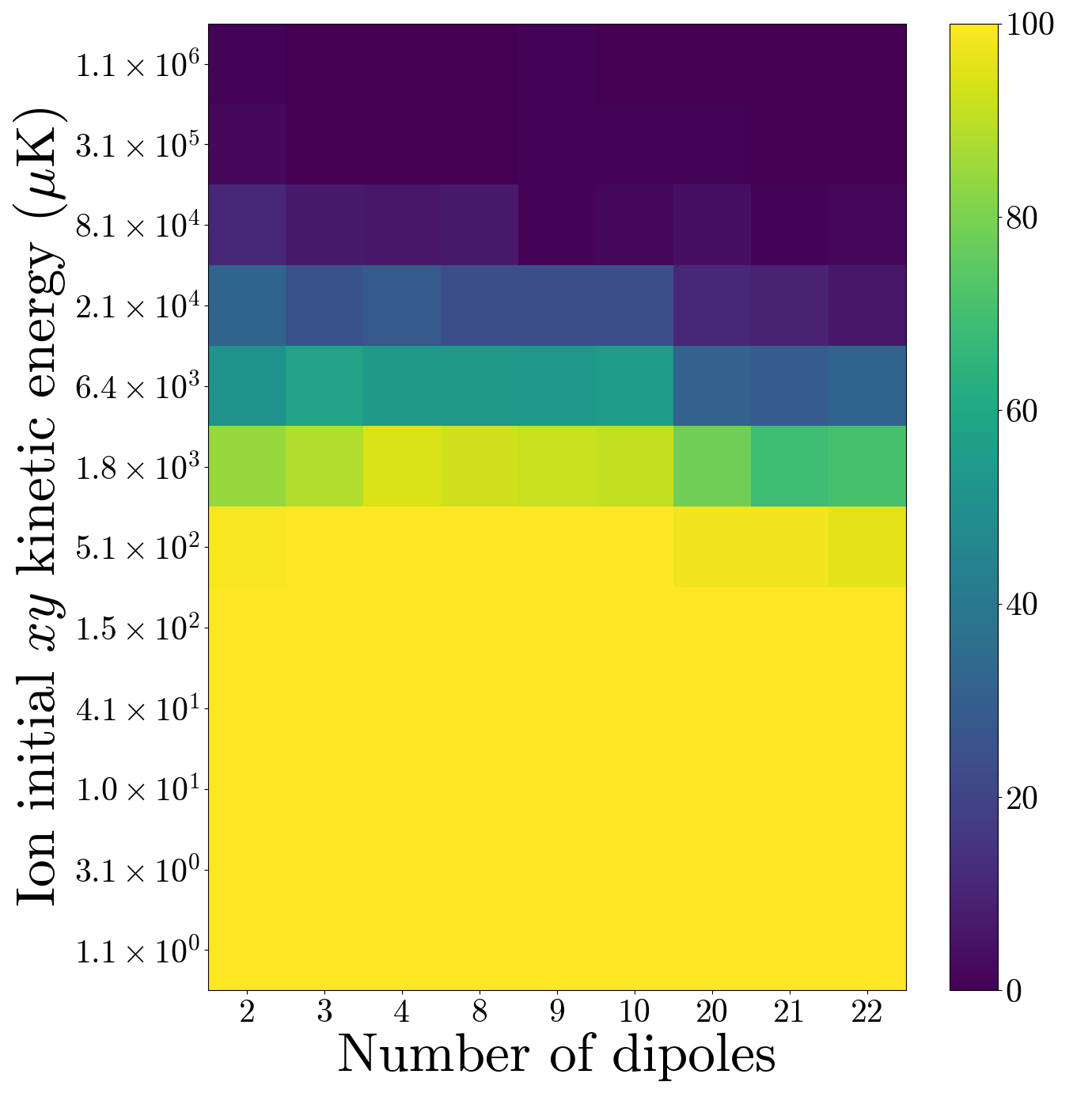}}
   \hfill
   \subfloat[]{\includegraphics[width=0.33\textwidth]{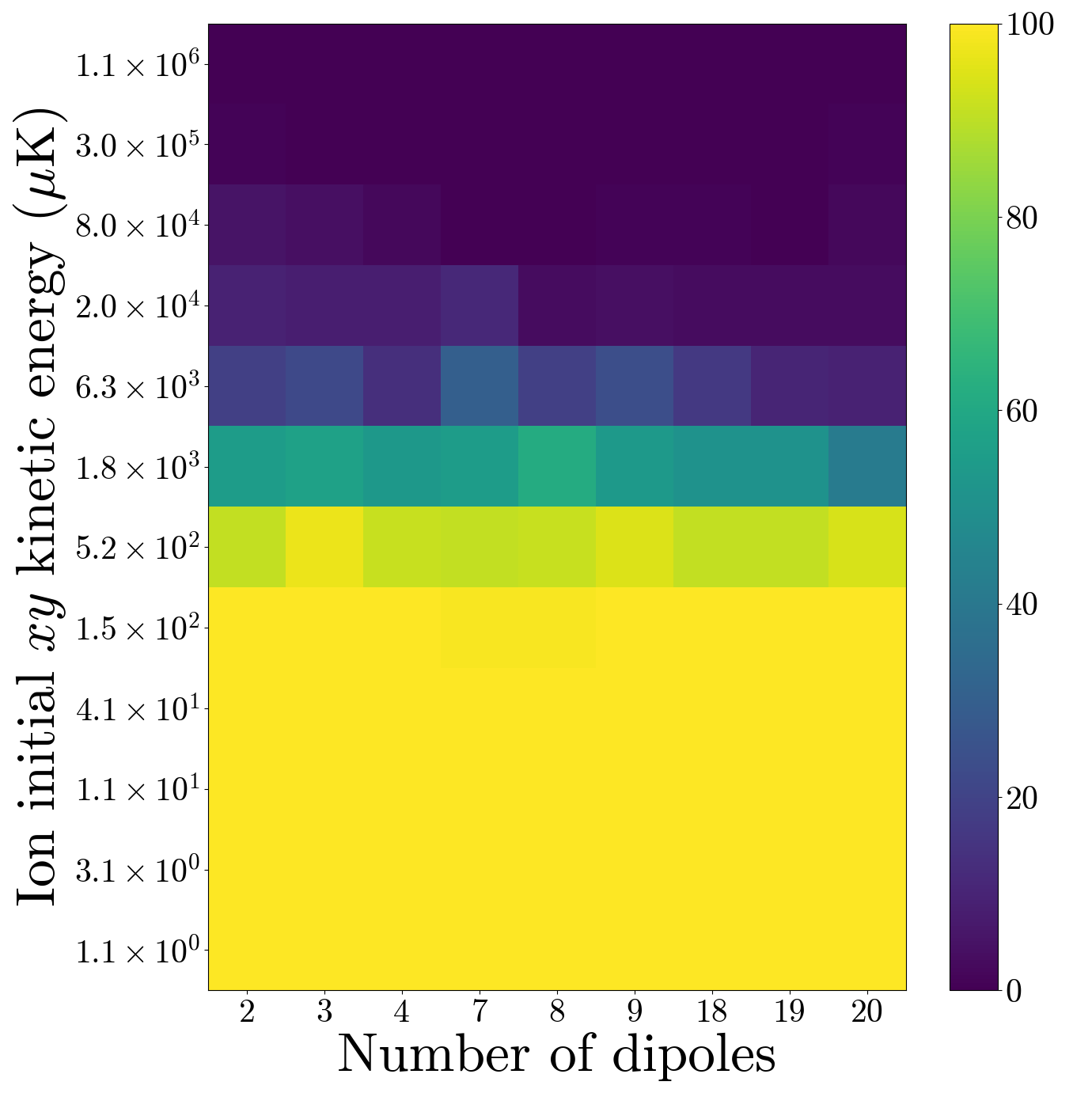}}
   \caption{Stability of ion–dipole clusters for different short-range barriers. Each panel shows the percentage of stable trajectories (out of 96 trajectories) as a function of the initial ion temperature in the $xy$ plane and the number of dipoles in the cluster. The initial ion temperature is calculated by averaging the kinetic energy of the ion in the x and y directions over two secular periods at the beginning of the simulation. The cluster sizes are chosen to span 2, 3, and 4, as well as values centered around the first and second magic numbers. The color scale indicates the percentage of stable trajectories, and the red curve denotes the evaporation energy as a function of cluster size. The $C_8^{\text{id}}$ and $C_8^{\text{dd}}$ for panels (a), (b), and (c), in atomic units, are $1 \times 10^{23}$ and $5 \times 10^{20}$, $1 \times 10^{23}$ and $8 \times 10^{20}$, and $5 \times 10^{22}$ and $8 \times 10^{20}$, respectively.} 
   \label{fig:stable}
\end{figure*}

By repeating this analysis while varying the number of molecules, the initial ion temperature, and the short-range interaction parameters, we obtain the stability diagram shown in Fig.~\ref{fig:stable}. A notable result is that an initial ion temperature of approximately $1\,\mathrm{mK}$ is required to dissociate clusters with evaporation energies of the order of $10\,\mu\mathrm{K}$. This seemingly counterintuitive difference arises because only a fraction of the ion's motional energy is transferred to a given molecule during an ion–dipole encounter. To eject a molecule, the ion must transfer at least the evaporation energy to that molecule. Moreover, the ion must move away from the trap center to approach a dipole, which requires energy determined by the ion mass and secular trapping frequency. The dissociation threshold is therefore controlled not only by the evaporation energy but also by the efficiency of ion-to-dipole energy transfer and by the energy required to displace the ion within the trap. Importantly, this threshold depends strongly on the local shape of the ion–dipole potential near its equilibrium separation, which is controlled by $C_8^{\mathrm{id}}$.

A second important observation is that the boundary between stable and unstable trajectories depends only weakly on the number of molecules in the cluster. This behavior can be understood from the predominantly local nature of the dissociation mechanism. Because the two-body ion–dipole potential is largely independent of the total number of dipoles, the energy required to initiate the evaporation of a single dipole varies only weakly with cluster size. This observation also explains why $C_8^{\mathrm{id}}$ has a much stronger influence on the cluster dynamics than the corresponding dipole–dipole coefficient, $C_8^{\mathrm{dd}}$.
Comparing Figs.~\ref{fig:stable}(a) and \ref{fig:stable}(b), in which $C_8^{\mathrm{dd}}$ is varied while $C_8^{\mathrm{id}}$ is held fixed, shows that the overall stability changes only slightly. In contrast, comparing Figs.~\ref{fig:stable}(b) and \ref{fig:stable}(c), in which $C_8^{\mathrm{id}}$ is reduced while $C_8^{\mathrm{dd}}$ remains constant, reveals that the clusters become unstable at substantially lower initial ion temperatures. This behavior can be understood from the ion–dipole potential. Reducing $C_8^{\mathrm{id}}$ shifts the equilibrium separation and modifies the curvature of the potential near its minimum. In particular, a given increase in the interaction energy above the minimum can be reached with a smaller change in the ion–dipole separation. Consequently, the ion must undergo a smaller displacement within the trap to access the strongly repulsive region of the potential, reducing the ion kinetic energy required to initiate molecular evaporation.

\begin{figure}
    \centering
    \includegraphics[width=\linewidth]{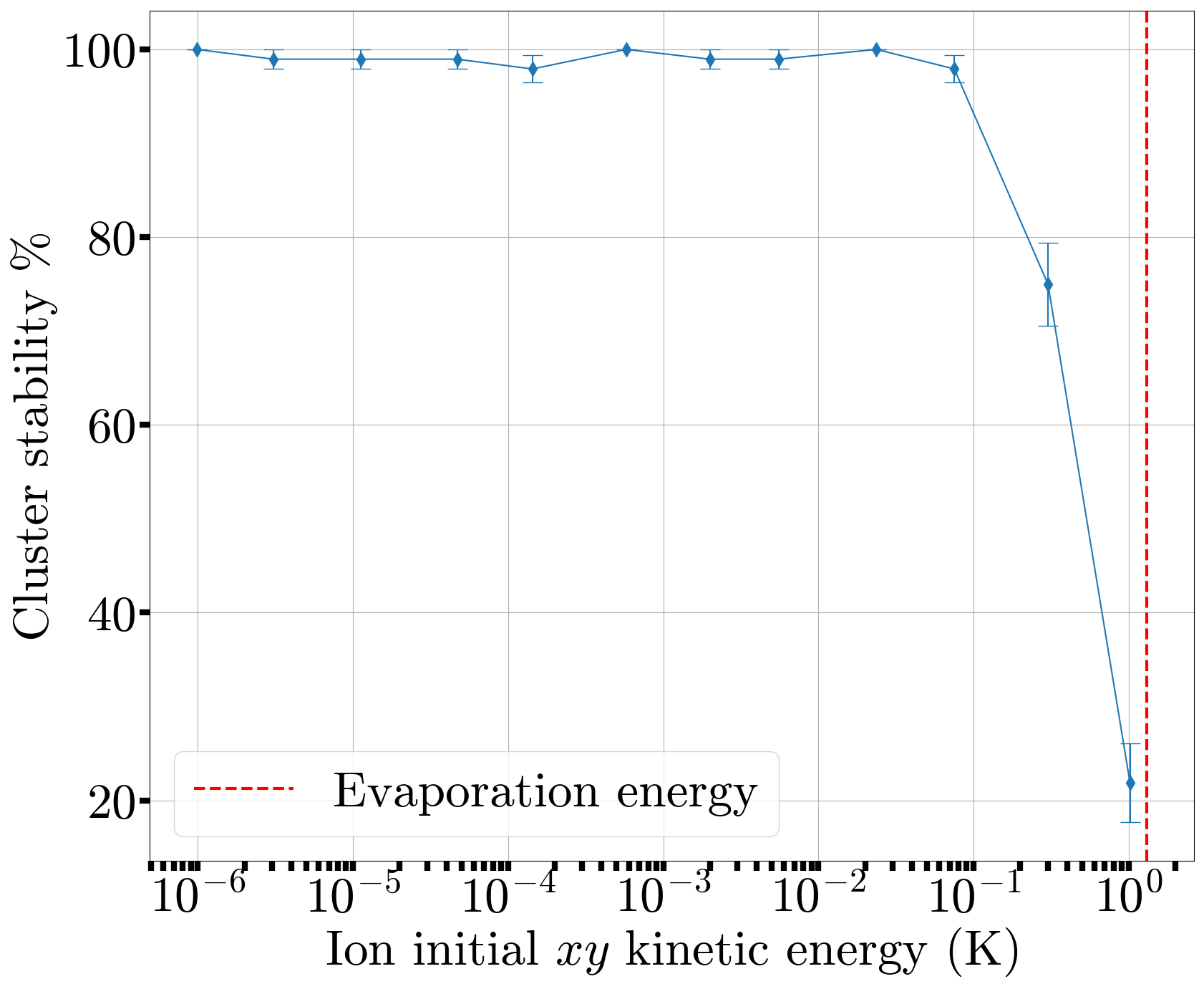}
    \caption{Stability of ion-dipole cluster around the ion as a function of ion's initial kinetic energy in $x$ and $y$ directions. The $C_8^{\text{id}}$ and $C_8^{\text{dd}}$ in atomic units are both $1\times10^{15}$. The cluster contains 5 dipoles as it is the first magic number of this configuration. The red dotted line represents the evaporation energy is 1.3K.}
    \label{fig:fig9}
\end{figure}

By contrast, when the system forms a crystalline molecular ring, the ion's initial kinetic energy below the evaporation energy is sufficient to destabilize the crystal, as shown in Fig.~\ref{fig:fig9}. For the strongly localized parameter set examined here, destabilization occurs at ion kinetic energies below the nominal evaporation energy, indicating that rigidity alone does not guarantee dynamical robustness.


\subsection{Towards 3D geometries}
We have extended our calculations to three-dimensional confinement. The basin-hopping evaporation energies as a function of the number of molecules are shown in Fig.~\ref{fig:fig9}. In analogy with the two-dimensional case, the molecules arrange themselves in concentric shells around the ion. Because of their larger available surface area, these spherical shells can accommodate more molecules than the corresponding rings in two dimensions.

As in the two-dimensional system, the magic numbers identify particularly stable configurations associated with the completion of successive coordination shells. In three dimensions, however, the evaporation energy varies nearly monotonically with the number of molecules after completion of the first shell. Consequently, the boundaries between the second and third shells are less clearly resolved. This behavior is similar to that observed for large two-dimensional clusters, as shown in Fig.~\ref{fig:fig5}.

We expect only the first shell to exhibit appreciable rigidity and a crystal-like structure. Molecules in the second and subsequent shells are expected to be substantially more delocalized, consistent with increasingly liquid-like behavior. As in two dimensions, the properties of the first shell may retain some dependence on the ion mass. This dependence should, however, rapidly diminish for the second and more distant shells.

\begin{figure}
    \includegraphics[width=\linewidth]{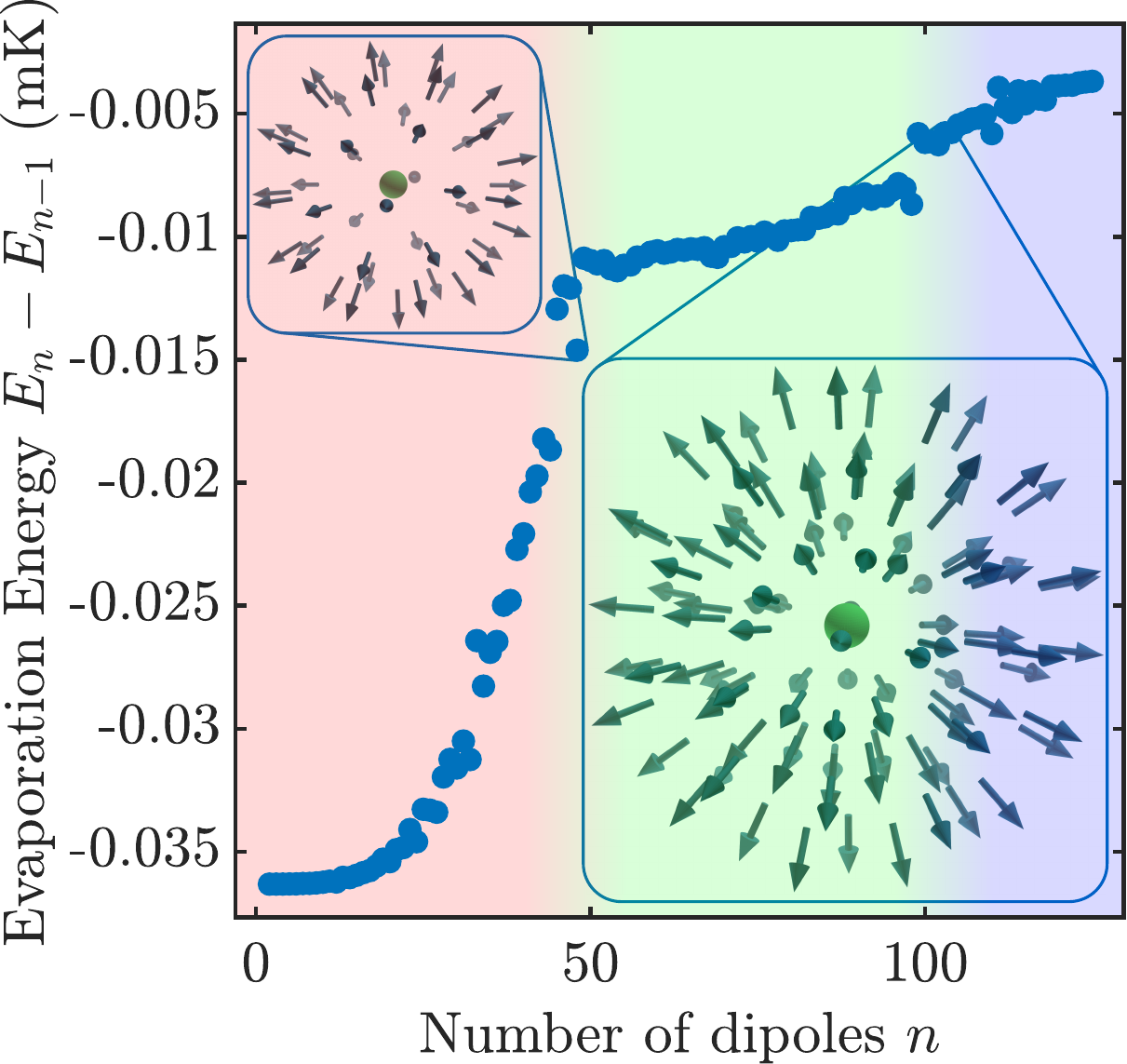}
    \caption{Evaporation energies of 3D dipolar clusters surrounding an ion. Differences between global minimum energies for 3-dimensional clusters of dipoles surrounding an ion computed using basin hopping, with the $C_8$ coefficients used in Fig.~\ref{fig:fig2}. The insets show the structures for the two magic numbers 48 and 103.}
    \label{fig:fig10}
\end{figure}

\section{Experimental considerations}
The described studies of ion-molecule interactions require a new hybrid approach combining methodology from the fields of ion-atom mixtures and molecular quantum gases, as proposed in \cite{Karpa2025}. The requirements for a suitable experimental apparatus are largely comparable to capabilities demonstrated recently in studies of ultracold ion-atom collisions \cite{Feldker2020,Schmidt2020,Weckesser2021}, including extended protocols for immersing ions into neutral ensembles using composite optical traps which avoid limitations arising from the presence of radiofrequency fields in conventional Paul traps \cite{Schmidt2020}.

In order to prepare quasi-2D ensembles of molecular quantum gases, recently developed techniques utilizing a combination of 1D optical lattices \cite{Valtolina2020} can be adopted for the scenario described in this work. For ensembles of $10^{3} - 10^{4}$ molecules prepared in typical experiments, we expect that realistic lattices with $10^{2} - 10^{3}$ molecules per layer can be achieved in an ion-molecule hybrid setup after implementing an additional 1D optical lattice.

\begin{figure}
	\includegraphics[width=\linewidth]{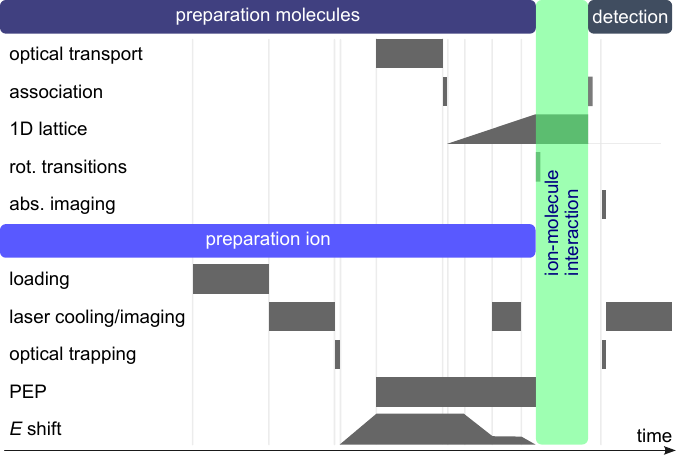}
	\caption{Schematic representation of a basic experimental protocol for combining an ion and molecular quantum gases. This scheme is based on methods developed for ultracold ion-atom experiments adopted for on-site association of molecules from a two-species mixture optically transported to the science chamber and subsequently loaded into a 1D optical lattice similar to \cite{Valtolina2020}. Crucial methods for observing ion-molecule interactions include optical trapping for the deterministic preparation of single ions and temperature measurements, and techniques for the efficient removal of parasitic ions using a controlled displacement of the ion with an electric offset field ($E$ shift) in combination with parametric excitation pulses (PEP) as reported in \cite{Schmidt2020}. The overall timescales are comparable to Ref.\cite{Schmidt2020} while the duration of individual steps depends on the characteristics of the apparatus. The ramp durations for the 1D lattice and $E_{shift}$ are long compared to the inverse trapping frequency, typically $1-10 ~ \textrm{ms}$, with ion loading and cooling periods on the order of $1-10 ~ \textrm{s}$, and $10-100 ~\textrm{ms}$ for laser pre-cooling before the interaction phase. Longer bars indicate durations that are much longer than the pulses for molecule association, rotational transitions, absorption imaging ($10-100 ~\mu\textrm{s}$), and optical ion trapping ($1 ~\textrm{ms}$).}
	\label{fig:protocol}
\end{figure}

A basic protocol for carrying out the proposed experiments involves a combination of the ion-atom overlapping scheme reported in \cite{Schmidt2020} and routinely employed methods for preparing ground-state molecules in a bulk and in a 1D optical lattice \cite{Valtolina2020}. To this end, the preparation of an atomic ensemble in a bulk trap is extended towards molecular quantum gases loaded into an optical lattice \cite{Valtolina2020} as illustrated in Fig. \ref{fig:protocol}. This can be accomplished by adopting a combination of established techniques. For the ion, this includes the deterministic preparation of a single ion \cite{Weckesser2021a}, laser (Doppler or Raman sideband) cooling, stray electric field compensation to the level of $E_\textrm{str} = 10^{-3} ~\textrm{V m}^{-1}$ \cite{Schmidt2020,Weckesser2021,Weckesser2021a}, and optical trapping \cite{Lambrecht2017}, followed by a controlled shift away from the loading region of the molecules and mass-selective removal of potentially created parasitic ions using parametric excitation pulses (PEP) as shown in previous works \cite{Schmidt2020}. An ensemble of molecules can be prepared using standard techniques by on-site association of two overlapped atomic clouds to loosely bound Feshbach molecules, followed by coherent two-photon transfer to the rovibronic ground state using a STIRAP (Stimulated Raman Adiabatic Passage) sequence. Subsequently, the optical dipole trap confining the molecules can be superimposed with an optical 1D standing wave, distributing the molecules among a defined number of lattice sites (on the order of 10) where the strong confinement along the $z$-axis ensures that the sub-ensembles are effectively confined to a 2D geometry. As a final step, the previously prepared ion can be controllably moved to a defined position away from a selected layer or placed into its center using electrostatic potentials. The required sub-wavelength resolution has been demonstrated in previous experiments \cite{Karpa2013}.

While a hybrid approach combining three species is substantially more complex compared to typical diatomic molecule or ion-atom experiments, it also provides crucial advantages making it suitable for the proposed studies. 

First, compared to ion-atom interactions, mixtures of ions and molecules offer access to and control over additional degrees of freedom, e.g., by exploiting the rotational structure of the molecules. This allows tuning the magnitude and sign of static polarizabilities $\alpha(E) = -\frac{\partial^2 U}{\partial E^2}$ and dipole moments $d(E) = -\frac{\partial U}{\partial E}$, by selectively preparing molecules in a specific state $\ket{J, m_J}$ using microwave fields. As an example, molecules in $\ket{J = 1, m_J = 0}$, a low-field seeking state with positive curvature with respect to $E$ as depicted in Fig. \ref{fig:fig1}(a), experience repulsive charge-induced dipole interactions, which can be used to initialize the molecules in a ring (for quasi-2D ensembles) or sphere-like (for bulk gases) distribution around the ion. This can be achieved by applying a $\pi$-pulse on the rotational transition $ \ket{J = 0, m_J = 0} \rightarrow \ket{J = 1, m_J = 0}$ at a frequency $\omega_J = 2B_0/\hbar$ (at $E=0$), with pulse durations on the order of $10 ~\mu \textrm{s}$ at typically available intensities of the microwave field. This enhances shot-to-shot reproducibility and at the same time prevents inelastic collisions during the preparation stage. %

Second, for molecules occupying a ring or a shell, ion-mediated shielding is expected to suppress undesired short-range collisions leading to inelastic losses \cite{Karpa2025}. This is a consequence of the repulsive intermolecular dipole-dipole interactions which also stabilize the mesoscopic molecular ion by counteracting the attractive charge-dipole forces between the ion and the molecules.

Third, combining ions and molecules affords access to additional observables signaling the formation of mesoscopic molecular ions. For example, the strong gradient of the electric field of the ion gives rise to sizeable relative shifts of the rotational transitions experienced by molecules within the cluster compared to unperturbed or surrounding molecules as indicated in Fig. \ref{fig:fig1}(a). This allows for high-contrast or background-free detection schemes where all molecules outside of the mesoscopic molecular ion or in neighboring sites of the 1D optical lattice are transferred to a specific rotationally excited state, e.g., within the $ \ket{J = 1}$ manifold, dissociated, and removed. In addition to detecting the presence of mesoscopic molecular ions, this approach can be extended to spectroscopically determine the ion-dipole equilibrium distance $r_{e,\textrm{id}}$, providing information about the size of the cluster. To this end, the molecules confined within a ring or shell can be transferred to the $\ket{J = 1, m_J = 0}$ state exhibiting repulsive charge-dipole interaction leading to ejection of the molecules if the microwave field is resonant with the $ \ket{J = 0, m_J = 0} \rightarrow \ket{J = 1, m_J = 0}$ transition at $r_{e,\textrm{id}}$.

Another principal advantage of this hybrid approach is the possibility to perform correlation measurements of ion and molecule observables. For instance, correlated changes of the effective masses of the molecules and ions can be probed by measuring the corresponding trap frequencies using resonant or parametric excitation. Another approach is to detect correlations between increased local molecular density and enhanced survival probabilities of the ion in the absence of radiofrequency confinement probing the expected localization by the dipolar cluster.

As an alternative approach to study mesoscopic molecular ions, the recently achieved control over individually trapped molecules allows preparing reconfigurable tweezer arrays \cite{Loic2019,applications,Guttridge2025}. In principle, this powerful technique can be used to initialize molecules in configurations considered in this study. This would enable experiments with a deterministically controlled number and spatial distribution of molecular dipoles.

\section{Conclusions and future work}
In this work, we have investigated the ground-state structure and stability of a single ion immersed in a two-dimensional ultracold gas of polar molecules. By calculating cluster energies as a function of the number of molecules bound to the ion, we obtained the corresponding evaporation energies and uncovered a characteristic plateau-like dependence on cluster size. These plateaus reflect the sequential formation of concentric molecular rings around the ionic impurity. The emergence of this ring structure is largely independent of the details of the short-range interactions, demonstrating that it originates from self-limiting electrostriction, due to the competition between the attractive ion–dipole interaction and dipole–dipole repulsion. The short-range potential nevertheless influences the number of molecules that can be accommodated within each ring. The rings of molecules around the ion form mesoscopic molecular ions with a rather flexible structure, but as the short-range ion-dipole repulsion weakens, the structures evolve into crystalline molecular rings.


We have also derived an approximate expression for the size and molecular capacity of the first ring under different confinement conditions. Because the derivation depends only weakly on the microscopic form of the short-range interaction, it provides a general framework for estimating the structure of ion-bound molecular clusters across a broad range of interaction parameters.

Finally, our dynamical stability analysis shows that mesoscopic molecular ions can survive initial ion temperatures substantially higher than the energy scale associated with molecular evaporation, whereas crystalline molecular rings cannot. The dissociation threshold depends only weakly on the total number of molecules and is governed primarily by the local shape of the ion–dipole potential and the efficiency of energy transfer from the trapped ion to an individual molecule. These findings indicate that ion-bound polar-molecule clusters should be stable under experimentally accessible conditions. More broadly, our results establish a distinct class of mesoscopic molecular ions whose structures are determined by self-limiting electrostriction. Under particular short-range ion-dipole interactions, these structures will evolve into crystalline molecular rings, opening a route to the study of charged impurities in quantum molecular environments.

\section{Acknowledgments}
S.C., R.S. and J.P.-R. acknowledge the support of the United States Air Force Office of Scientific Research [grant number FA9550-23-1-0202]. L.K. thanks the Deutsche Forschungsgemeinschaft (DFG, German Research Foundation) for support through the Heisenberg Programme No. 506287139 and under Germany’s Excellence Strategy – EXC 2123/2 QuantumFrontiers – 390837967.

\appendix 
\section{BH convergence}
To find the global minimum, basin hopping does a Monte Carlo Metropolis sampling of local minima with gradient descent, and this Monte Carlo is more effective when there are more independent parallel cores and when the number of trials is increased. 


One way to characterize convergence is to plot the evaporation energies as a function of $n$, for several different numbers of iterations (or several different wall clock times), and observe the convergence of the plot with respect to the wall clock time. If this plot, for the range of $n$'s in this study, does not differ between the wall time $t$ that we used and wall times $t/2$ and $2t$, then the basin hopping has very likely converged to the global minimum. 

We have two different basin hopping techniques, one basic BH technique used exclusively for 2d clusters in the low-density regime, and another more advanced BH technique used for the 3D clusters and higher-density regime 2d clusters. Thus, for the $C_8$ combinations shown in Fig.~\ref{fig:fig2} and Fig.~\ref{fig:fig9}, we choose 10 different iteration numbers and plot the evaporation energies as a function of $n$. The results are shown in Fig.~\ref{fig:fig10}. For $n\leq 12$, even just 10 iterations of basin hopping are enough to converge to the putative global minimum.


\begin{figure}[h]
    \includegraphics[width=\linewidth]{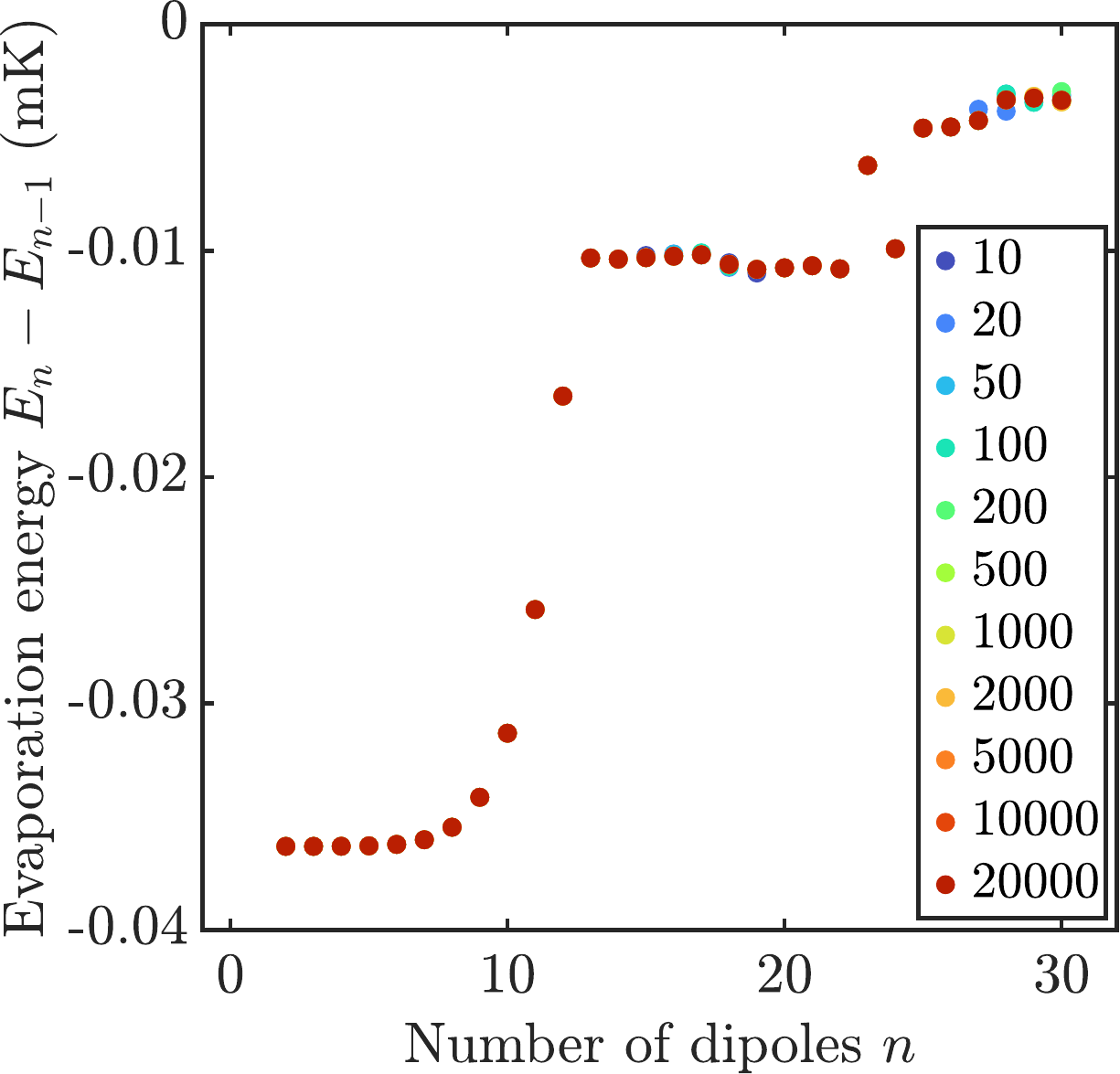}
    \caption{Evaporation energies calculated by the basic basin hopping technique for the same potential as in Fig.~\ref{fig:fig2}, for different numbers of basin hopping iterations shown in the legend, parallelized over 28 cores, with the same set of random seeds for each number of iterations.}
    \label{fig:fig10}
\end{figure}

\section{DMC convergence}
\begin{figure}[h]
    \includegraphics[width=\linewidth]{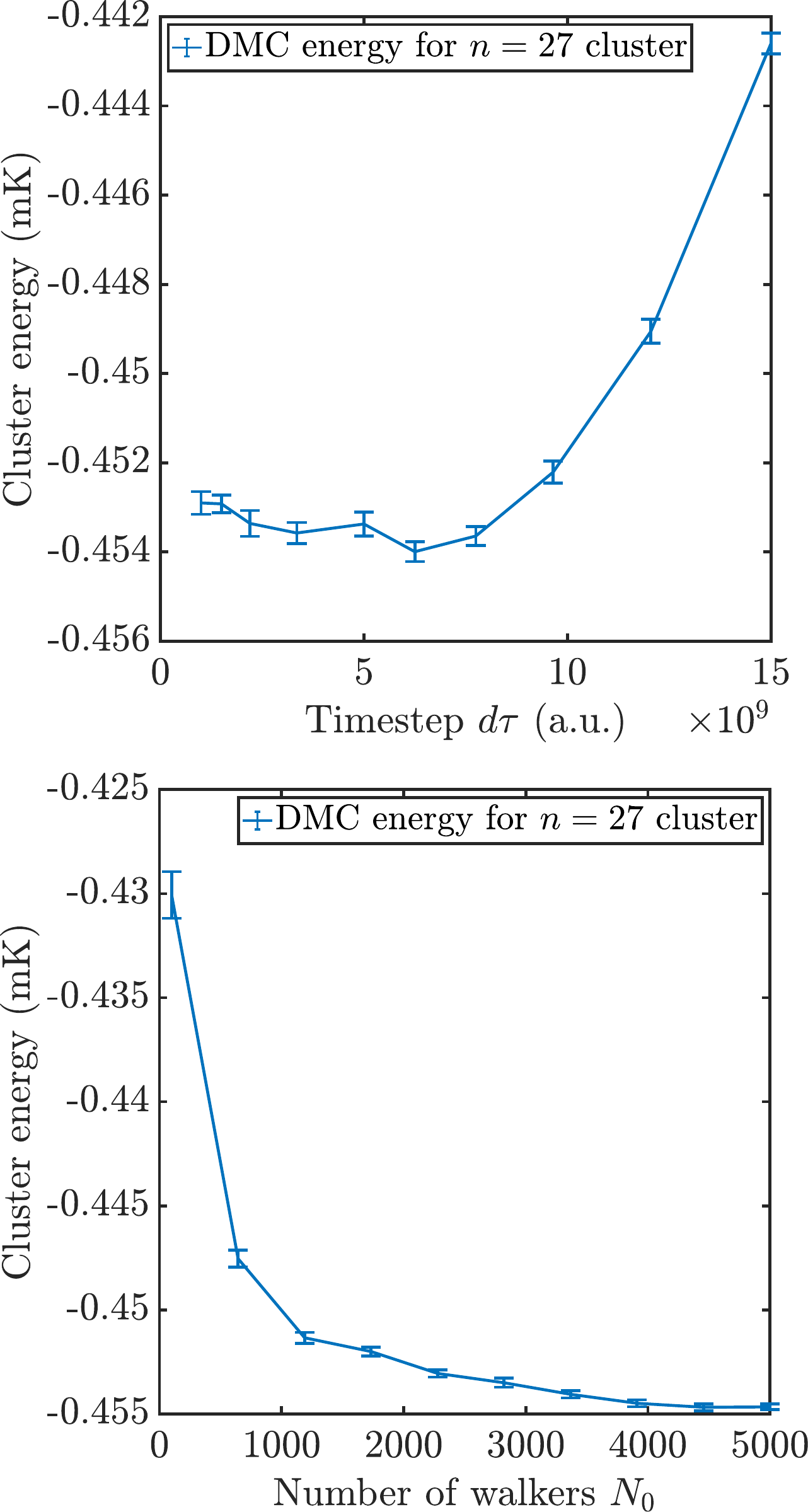}
    \caption{DMC convergence tests for the second magic number $n=27$ cluster for the potential given by $C_{8,\textrm{id}}=1\times 10^{22}$ and $C_{8,\textrm{dd}}=5\times 10^{18}$, which is the same potential as in Fig.~\ref{fig:fig2}.}
    \label{fig:fig11}
\end{figure}

DMC is a numerical technique which uses a discrete infinitesimal imaginary timestep $d\tau$ and a finite number of walkers $N_0$. In theory, in the limit as $d\tau\to 0$ and $N_0\to\infty$, the DMC energy converges to the true ground state energy of the system. Thus, to ensure convergence, we pick a combination of $C_8$'s in the low density regime and check both of these limits for a particular cluster. As shown in Fig.~\ref{fig:fig11}, our DMC results are converged. In the main text, we used $d\tau = 5\times 10^{9}$ a.u. and $N_0=2000$ for this $C_8$ combination. If we had used 5000 walkers instead of 2000, the energy for the $n=27$ cluster would have changed by about 0.5\%. If we had used a timestep increment of $1\times 10^{9}$ a.u. instead of $5\times 10^{9}$ a.u., the energy would have changed by about 0.5\% as well. 

\section{Magic number derivation}

As discussed in the main text, the first magic number geometry is roughly a circular ring of $n$ dipoles surrounding the ion at radius $r_e$, the equilibrium distance of the ion-dipole interaction. The $i^\textrm{th}$ dipole in this ring is located at $\vec{r}_i=(r_e\cos\frac{2\pi i}{n}, r_e\sin\frac{2\pi i}{n}, 0)$. It is useful to find the dot products of $\vec{r}_i$, $\vec{r}_j$, $\vec{r}_{ij}=\vec{r}_j-\vec{r}_i$, and the magnitude of $\vec{r}_{ij}$. We find that $\vec{r}_i\cdot\vec{r}_j=r_e^2\cos\frac{2\pi(i-j)}{n}$, $r_{ij}=2r_e\left|\sin\frac{\pi(i-j)}{n}\right|$, and $\vec{r}_i\cdot\vec{r}_{ij}=r_e^2\left(\cos\frac{2\pi(i-j)}{n}-1\right)$.

Substituting these into Eq.~\ref{Total_pot}, we obtain the potential energy of the ring,
\begin{multline}
    V_\textrm{ring}(n)=n\left(\frac{C_8^\textrm{id}}{r_e^8}+U\left(\frac{1}{r_e^2}\right)\right)+\sum\limits_{i=1}^{n-1}\sum_{j=i+1}^{n}\Bigg(\\
    \frac{C_8^\textrm{dd}}{r_e^8\left(2\sin\frac{\pi(i-j)}{n}\right)^8}+\frac{d(r_e)^2}{r_e^3\left|2\sin\frac{\pi(i-j)}{n}\right|^3}\cdot\\\frac{1}{r_e^2}\Bigg(
    r_e^2\cos\frac{2\pi(i-j)}{n}+\frac{3r_e^4\left(1-\cos\frac{2\pi(i-j)}{n}\right)^2}{r_e^2\left(2\sin\frac{\pi(i-j)}{n}\right)^2}\Bigg)\Bigg).
\end{multline}
The last line above can be simplified to $\cos\frac{2\pi(i-j)}{n}+3\sin^2\frac{\pi(i-j)}{n}$. The ion-dipole potential at distance $r_e$ has value $-D_e$. The double summation is a sum, over all possible distances between two vertices of a regular polygon, of some function of the distance. There are only $\left\lfloor\frac{n}{2}\right\rfloor$ unique such distances. If we index the vertices as $i=1,2,\ldots,n$, then the summation can also be thought of as a sum over all index differences $k=|i-j|$. If we count the distance-dependent function $n/2$ times for each index-difference $k=1,\ldots,n-1$, then we would correctly count all non-main-diagonal distances $n$ times (since index differences $k$ and $n-k$ both yield the same distance), and count main-diagonal distances $n/2$ times (since $k=n/2$ is counted only once, only when $n$ is even). Thus, the double-summation can be rewritten as a single summation
\begin{multline}
    V_\textrm{ring}=-nD_e+\\
    \frac{n}{2}\sum\limits_{k=1}^{n-1}\left(\frac{C_8^\textrm{dd}}{r_e^8}\frac{1}{\left(2\sin\frac{\pi k}{n}\right)^8}+\frac{d(r_e)^2}{r_e^3}\frac{1+\sin^2\frac{\pi k}{n}}{\left(2\sin\frac{\pi k}{n}\right)^3}\right).
\end{multline}
It can be verified algebraically with geometric sequence summation that $\sum\limits_{k=1}^{n-1}\csc^8\frac{\pi k}{n}=\frac{1}{14175}\left(3n^8+40n^6+294n^4+2160n^2-2497\right)$. According to Theorem 4b in \cite{Blagouchine2024FiniteCosecants}, $\sum\limits_{k=1}^{n-1}\csc\frac{\pi k}{n}\approx \frac{2n}{\pi}\left(\log\frac{2n}{\pi}+\gamma\right)-\frac{\pi}{36 n}$, where $\gamma$ is the Euler-Mascheroni constant. Finally, using the Mittag-Leffler expansion of $\csc^3x$ and Euler-MacLaurin expansions of sums of signed $r^{-3}$ terms, it can be shown that $\sum\limits_{k=1}^{n-1}\csc^3\frac{\pi k}{n}\approx\frac{2\zeta(3)}{\pi^3}n^3+\frac{n}{\pi}\left(\log\frac{2n}{\pi}+\gamma\right)-\frac{n}{6\pi}-\frac{17\pi}{720n}$, where $\zeta(z)$ is the Riemann zeta function. Substituting these approximations, we obtain
\begin{multline}\label{eq:Vringappendix}
    V_\textrm{ring}(n)\approx -nD_e\\
    +n^9\frac{C_8^\textrm{dd}}{r_e^8}\frac{1}{512\cdot 14175}\Bigg(3+\frac{40}{n^2}+\frac{294}{n^4}+\frac{2160}{n^6}-\frac{2497}{n^8}\Bigg)\\
    +n^4\frac{d(r_e)^2}{r_e^3}\frac{1}{16}\Bigg(\frac{2\zeta(3)}{\pi^3}+\frac{3}{\pi n^2}\left(\log\frac{2n}{\pi}+\gamma\right)\\
    -\frac{1}{6\pi n^2}-\frac{37\pi}{720n^4}\Bigg).
\end{multline}

If $n$ is the first magic number, then another dipole would join outside the ring at position $\vec{r}_x=(x\cos\frac{\pi}{n},x\sin\frac{\pi}{n},0)$ next to two dipoles within the ring, forming an isosceles triangle which we observed to often be nearly equilateral. We begin by computing vector dot products and magnitudes: $\vec{r}_i\cdot\vec{r}_x=r_ex\cos\frac{(2i-1)\pi}{n}$, $r_{ix}=r_e\sqrt{1+\frac{x^2}{r_e^2}-2\frac{x}{r_e}\cos\frac{(2i-1)\pi}{n}}$, $\vec{r}_i\cdot\vec{r}_{ix}=r_e^2\left(\frac{x}{r_e}\cos\frac{(2i-1)\pi}{n}-1\right)$, and $\vec{r}_x\cdot\vec{r}_{ix}=r_e^2\left(\frac{x^2}{r_e^2}-\frac{x}{r_e}\cos\frac{(2i-1)\pi}{n}\right)$. Substituting these expressions into the dipole-dipole interaction expression and summing over the $n$ dipoles, and allowing the outer dipole to relax by minimizing over possible ion-dipole distances $x$, the new energy added by the outer dipole is
\begin{multline}\label{eq:Enew}
    E_\textrm{new}(n)=\min_x\Bigg(\frac{C_8^\textrm{id}}{x^8}+U\left(\frac{1}{x^2}\right)\\
    +\sum\limits_{i=1}^n\Bigg(\frac{C_8^\textrm{dd}}{r_e^8\left(1+\frac{x^2}{r_e^2}-\frac{2x}{r_e}\cos\frac{(2i-1)\pi}{n}\right)^4}\\
    +\frac{d(r_e)d(x)}{r_e^3\left(1+\frac{x^2}{r_e^2}-\frac{2x}{r_e}\cos\frac{(2i-1)\pi}{n}\right)^{3/2}}\Bigg(\cos\frac{(2i-1)\pi}{n}\\
    -\frac{3r_e}{x}\frac{\left(\frac{x}{r_e}\cos\frac{(2i-1)\pi}{n}-1\right)\left(\frac{x^2}{r_e^2}-\frac{x}{r_e}\cos\frac{(2i-1)\pi}{n}\right)}{1+\frac{x^2}{r_e^2}-\frac{2x}{r_e}\cos\frac{(2i-1)\pi}{n}}\Bigg)\Bigg)\Bigg).
\end{multline}


As discussed in the main text, the first magic number is the smallest $n$ such that $V_\textrm{ring}(n)+E_\textrm{new}(n)<V_\textrm{ring}(n+1)$. We can make a number of approximations to simplify this model. First, since the electric field at around 10000$a_0$ is weak, $U(r)\approx -\frac{1}{2}\alpha E(r)^2=-\frac{C_4}{r^4}$, where $C_4=\alpha/2$ in atomic units and $\alpha$ is the polarizability at weak fields. This implies that $r_e\approx \sqrt[4]{\frac{2C_8^\textrm{id}}{C_4}}$. Second, our 2d clusters consistently follow the trend that the $(n+1)^\textrm{th}$ dipole forms roughly an equilateral triangle with the two nearest dipoles within the ring. Since the nearest-neighbor dipole spacing in the ring is roughly $2\pi r_e/n$, we have that $x\approx r_e+\frac{\sqrt{3}\pi r_e}{n}$ in Eq.~\ref{eq:Enew}. These two approximations enable us to avoid having to numerically solve an equation or optimize a function, thus enabling much quicker yet still accurate predictions by our model of the first magic numbers.

%


\end{document}